**Title:** Annotation-free deep learning for detection and segmentation of fetal germinal matrix-intraventricular hemorrhage on brain MRI

**Authors:** Mingxuan Liu[1,2#], BEng; Yingqi Hao[2#], BEng; Yi Liao[1#], MD; Juncheng Zhu[1], BSc; Haoxiang Li[2], BEng; Hongjia Yang[2], BEng; Yifei Chen[2], BEng; Yijin Li[2], BEng; Kasidit Anmahapong[2], BMed; Zihan Li[2], BEng; Jialan Zheng[2], BEng; Min Kang[3], MD; Yan Song[3], MD; Hua Lai[4], MD; Xiaoling Zhou[4], MD; Nan Sun[5], MMed; Rong Hu[6], MMed; Gang Ning[1], MD; Haibo Qu[1], MD; Qiyuan Tian[2], PhD.

[#]These authors contributed equally to this work.

**Author Affiliations:**

[1]Department of Radiology, West China Second University Hospital, Sichuan University, Chengdu, China;

[2]School of Biomedical Engineering, Tsinghua Medicine, Tsinghua University, Beijing, China;

[3]Department of Radiology, Sichuan Provincial Women's and Children's Hospital, The Affiliated Women's and Children's Hospital of Chengdu Medical College, Chengdu, China;

[4]Chengdu Women's and Children's Central Hospital, School of Medicine, University of Electronic Science and Technology of China, Chengdu, China;

[5]Department of Radiology, The Third Affiliated Hospital of Zhengzhou University, Zhengzhou, China;

[6]Qujing Maternal and Child Health Hospital, Qujing, China.

**Corresponding Author:** Qiyuan Tian, PhD, Center for Biomedical Imaging Research, Tsinghua University 30 Shuangqing Road, Haidian District, Beijing, China, PA100084 (qiyuantian@tsinghua.edu.cn). Haibo Qu, MD, No.20, Section 3, Renmin South Road, Chengdu, Sichuan Province, China, PA610041(windowsqhb@126.com).

**Abstract**

Prenatal germinal matrix-intraventricular hemorrhage (GMH-IVH) is a leading cause of infant mortality and neurodevelopmental impairment, yet its manual diagnosis and lesion segmentation on fetal brain MRI are labor-intensive and error-prone. Although supervised deep learning offers potential for automation, it typically requires large amounts of annotated GMH-IVH data, which are challenging to obtain for such a rare condition (0.5–0.9 per 1000 pregnancies). To address these problems, an annotation-free deep learning framework, FreeHemoSeg, was developed for automated detection and segmentation of GMH-IVH without any real patient annotations. Instead of learning from expert labels, FreeHemoSeg was trained on pseudo GMH-IVH images synthesized from normal fetal data guided by medical priors. The framework was evaluated in a retrospective multicentre study of 1,674 stacks of 2D $T_2$-weighted MRI from 558 pregnant women, using data from one hospital for internal training and validation and two hospitals for external validation. FreeHemoSeg achieved the highest diagnostic and segmentation performance in both internal validation (AUROC: 0.959; AUPR: 0.928; sensitivity: 0.914; specificity: 0.966; DSC: 0.559) and external validation (AUROC: 0.930; AUPR: 0.884; sensitivity: 0.824; specificity: 0.943; DSC: 0.512), outperforming a supervised model trained on limited empirical data and unsupervised anomaly detection methods. Moreover, FreeHemoSeg assistance improved radiologists' sensitivity (from 0.882 to 0.941–1.000) and diagnostic confidence, while reducing interpretation time by 16.0–52.7%. We anticipate its immediate utility in supporting earlier diagnosis, prognostic counselling, and perinatal planning for fetal GMH-IVH. Code: https://github.com/Arktis2022/FreeHemoSeg.

## 1. Introduction

Fetal germinal matrix-intraventricular hemorrhage (GMH-IVH) is the most common type of fetal brain hemorrhage and carries great clinical significance (Hao et al., 2026; Sanapo et al., 2017). The germinal matrix is a transient periventricular region comprising fragile thin-walled vessels and migrating neuronal components (Brouwer et al., 2014). During fetal development, neurons and glial cells migrate radially outward from the germinal matrix to the cerebral cortex. Due to the extreme fragility of the vasculature within the germinal matrix, various intrauterine insults (e.g., maternal conditions, placental complications, fetal alloimmune thrombocytopenia, and underlying genetic mutations) can lead to spontaneous hemorrhage in the germinal matrix that may extend into the lateral ventricles, constituting GMH-IVH (Hadi et al., 2024). Although fetal GMH-IVH frequently arises in association with maternal, placental, or fetal conditions rather than as an isolated event, it is consistently associated with adverse postnatal outcomes for the affected child, such as seizure disorders, intellectual disability, psychomotor delays, and cerebral palsy (Dunbar et al., 2021; Elchalal et al., 2005; Kim et al., 2024). Research suggests that prenatal localization and volume estimation of the hemorrhage could facilitate prenatal counseling and guide early referrals to suitable services, particularly when interpreted together with its underlying etiology, associated parenchymal injury, ventricular distortion, and concomitant developmental abnormalities (Dunbar et al., 2021; Elchalal et al., 2005). Therefore, the timely diagnosis of fetal GMH-IVH is of critical importance.

Magnetic resonance imaging (MRI) serves as a valuable complement to expert neurosonography, including transvaginal ultrasonography with multiplanar assessment, for diagnosing GMH-IVH (K.N. Epstein et al., 2021; Liu et al., 2026; Sanapo et al., 2017). MRI provides sequence-specific characterization of hemorrhage and associated injury, including low-signal blood products on $T_2$-weighted images, blooming from deoxyhemoglobin or hemosiderin on $T_2^*$-weighted or echo-planar imaging, and diffusion restriction related to acute ischemic injury on diffusion-weighted imaging (Moradi et al., 2024). However, diagnosing GMH-IVH and localizing hemorrhages still relies on subjective radiologist interpretation, which may overlook subtle lesions. In $T_2$-weighted MRI, hemorrhages and the germinal matrix both exhibit low signals, making them hard to distinguish. Although comprehensive multi-sequence protocols usually allow GMH-IVH to be differentiated from mimics such as gray matter heterotopia, this distinction becomes difficult when only $T_2$-weighted images are diagnostic. This occurs, for example, when supplementary $T_1$- or $T_2^*$-weighted sequences are unavailable or severely degraded by fetal motion. Furthermore, fetal motion during thick-slice imaging can introduce artifacts that obscure hemorrhages, hindering accurate assessment of severity and potentially delaying diagnosis and treatment. As a result, developing an automated and accurate diagnostic method for MRI is essential.

Deep learning holds great promise for accurate and automated segmentation of GMH-IVH lesions, as it has already been successfully applied to the segmentation and diagnosis of various brain lesions (Wang et al., 2025b, 2025a). In supervised segmentation, models are trained to establish a mapping from input images to voxel-wise labels, relying on large collections of expert-annotated masks (Bai et al., 2025; Huang et al., 2020; Li et al., 2018). However, the scalability of supervised learning for GMH-IVH is constrained by the limited availability of fetal cases (estimated at 0.5–0.9 per 1000 pregnancies) (Alessandro Parodi and Ramenghi, 2015) and the reliance on manual annotations, which are time-consuming, error-prone, and subject to inter- and intra-observer variability. To address the challenge of limited training data, annotation-free deep learning approaches have emerged (Chen, 2017). Conventional unsupervised anomaly detection (UAD) methods often suffer from limited accuracy (Liu et al., 2024; Zhou et al., 2022). Alternatively, pseudo-anomaly synthesis generates synthetic lesions from normal data using prior knowledge, showing promise for automated lesion detection and diagnosis (Chalcroft et al., 2024; Lyu et al., 2024; Yao et al., 2021; Zhang et al., 2024, 2023; Zong et al., 2026), but has not been applied to fetal GMH-IVH.

In this study, we propose FreeHemoSeg, an annotation-free deep learning framework based on pseudo-anomaly synthesis, to interpret fetal MRI scans and achieve accurate GMH-IVH diagnosis and segmentation. Intuitively, FreeHemoSeg works by synthesizing realistic pseudo GMH-IVH slices from normal fetal brain images under the guidance of medical priors (i.e., the periventricular origin of the germinal matrix and the characteristic $T_2$-weighted hypointensity of blood products graded by the Papile classification). These synthetic data are subsequently used to train a segmentation model for coarse lesion localization and to fine-tune a Segment Anything Model (SAM) for refined segmentation. The performance of FreeHemoSeg was systematically quantified in a retrospective multicentre study across three hospitals, in terms of case-level and slice-level diagnostic accuracy, as well as lesion segmentation accuracy. Furthermore, it was benchmarked against state-of-the-art (SOTA) UAD methods and a supervised model trained on scarce empirical data. We further evaluated its clinical utility through a reader study assessing radiologists' diagnostic accuracy, efficiency, and confidence. By removing the dependence on annotated fetal lesion data, FreeHemoSeg offers a practical route around the data scarcity that has long hindered automated fetal GMH-IVH assessment, with the potential to support earlier diagnosis, prognostic counselling, and perinatal planning.

**Table 1. Fetal MRI Acquisition Protocols for Different Scanners.** Abbreviations: HASTE, half-Fourier acquired single-shot turbo spin-echo; SSFSE, single shot fast spin echo; TSE-SSH, turbo spin-echo single-shot.

| Scanner Type | Model | Receive Coils | Sequence Type | Repetition Time (TR) | Echo Time (TE) | In-plane Resolution | Section Thickness | Acquisition Matrix |
|---|---|---|---|---|---|---|---|---|
| **Training Dataset and Internal Validation Dataset** | | | | | | | | |
| 3.0-T | Siemens MAGNETOM Skyra | Spine 18-channel body | HASTE | 1800-1900 msec | 80-100 msec | 1 × 1 mm² | 4 mm | 320 × 320 |
| 1.5-T | United Imaging uMR 570 | Spine 16-channel body | SSFSE | 1800-2000 msec | 90–110 msec | 1 × 1 mm² | 4 mm | 320 × 288 |
| 1.5-T | Philips Achieva | Spine 16-channel body | TSE-SSH | 1200–1400 msec | 80–120 msec | 1 × 1 mm² | 4 mm | 320 × 320 |
| **External Validation Dataset** | | | | | | | | |
| 1.5-T | GE Signa HDxt | 8-channel body | SSFSE | 1800 msec | 80-110 msec | 1 × 1 mm² | 4 mm | 512 × 320 |
| 3.0-T | Philips Ingenia | 16-channel body | TSE-SSH | 900-1500 msec | 90-130 msec | 1 × 1 mm² | 5 mm | 364 × 277 |

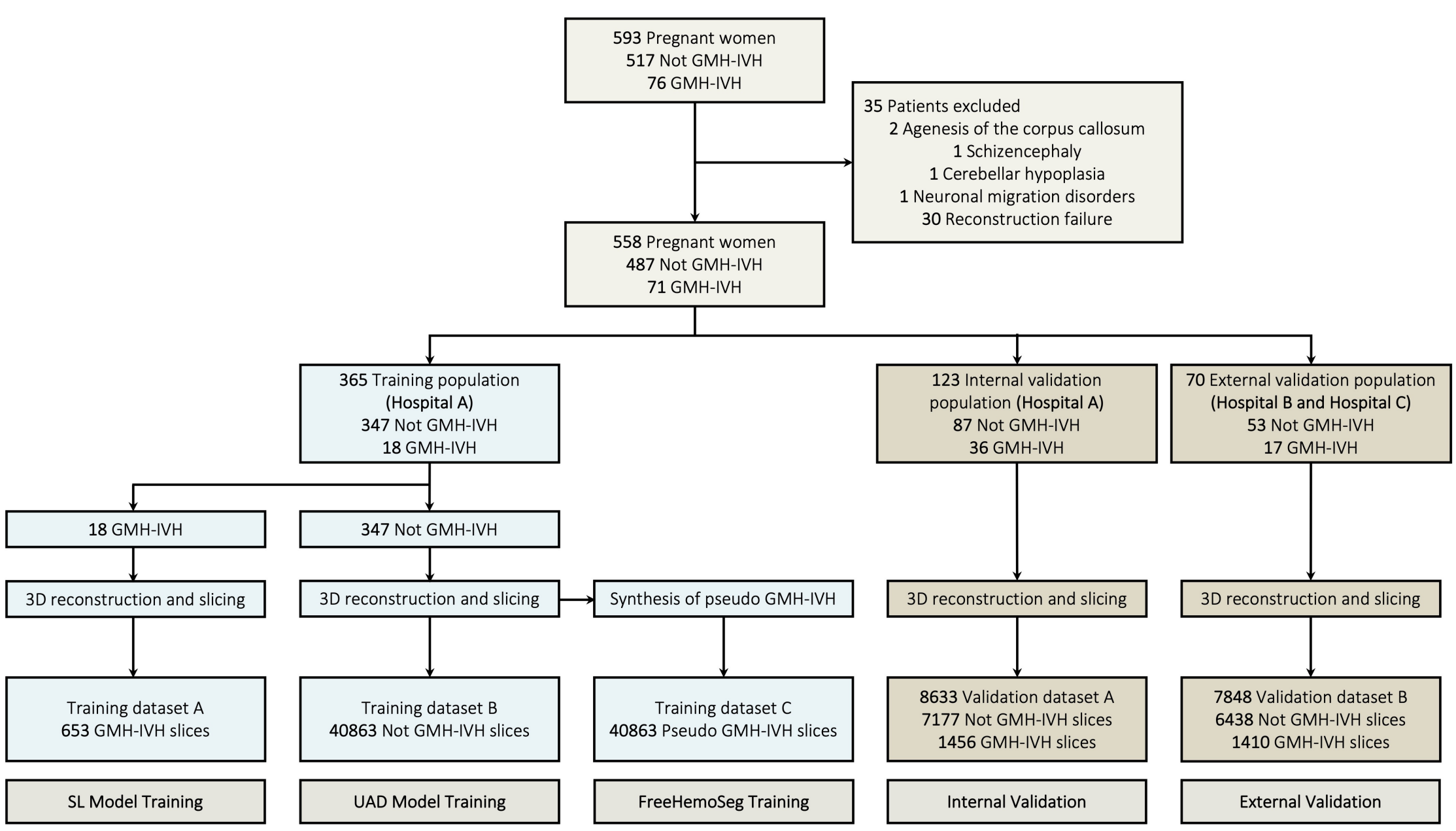

**Figure 1. Flow Diagram of Training and Validation Datasets.** Abbreviations: SL Model, supervised learning model; UAD, unsupervised anomaly detection; GMH-IVH, germinal matrix-intraventricular hemorrhage.

## 2. Materials and Methods

For this diagnostic study, the institutional review boards (IRBs) of Hospital A (West China Second University Hospital [WCSUH] of Sichuan University [NO. 2025-432]), Hospital B (Chengdu Women's and Children's Central Hospital [CWCCH; NO. 2026-7-2]), and Hospital C (Sichuan Provincial Woman's and Children's Hospital [SPWCH; NO. 20250122-010]) approved the retrospective use and analyses of fetal MRI data. Because of the retrospective design of the study, the requirement for informed consent was waived. This study is reported in accordance with the Standards for Reporting of Diagnostic Accuracy (STARD) 2015 guidelines, and adhered to the tenets of the Declaration of Helsinki.

### 2.1 Data Collection

To minimize image artifacts caused by fetal motion, 2D imaging sequences were employed to acquire T2-weighted MR images (Table 1). On the 3.0-T Siemens MAGNETOM Skyra scanner (WCSUH), a half-Fourier acquired single-shot turbo spin-echo (HASTE) sequence was used. On the 1.5-T United Imaging uMR 570 scanner (WCSUH) and 1.5-T GE Signa HDxt scanner (SPWCH), a single shot fast spin echo (SSFSE) sequence was used. On the 1.5-T Philips

Achieva scanner (WCSUH) and 3.0-T Philips Ingenia scanner (CWCCH), a turbo spin-echo single-shot (TSE-SSH) sequence was used. Imaging section thickness was 4–5 mm with no inter-slice gap. Image in-plane spatial resolution was 1 × 1 mm².

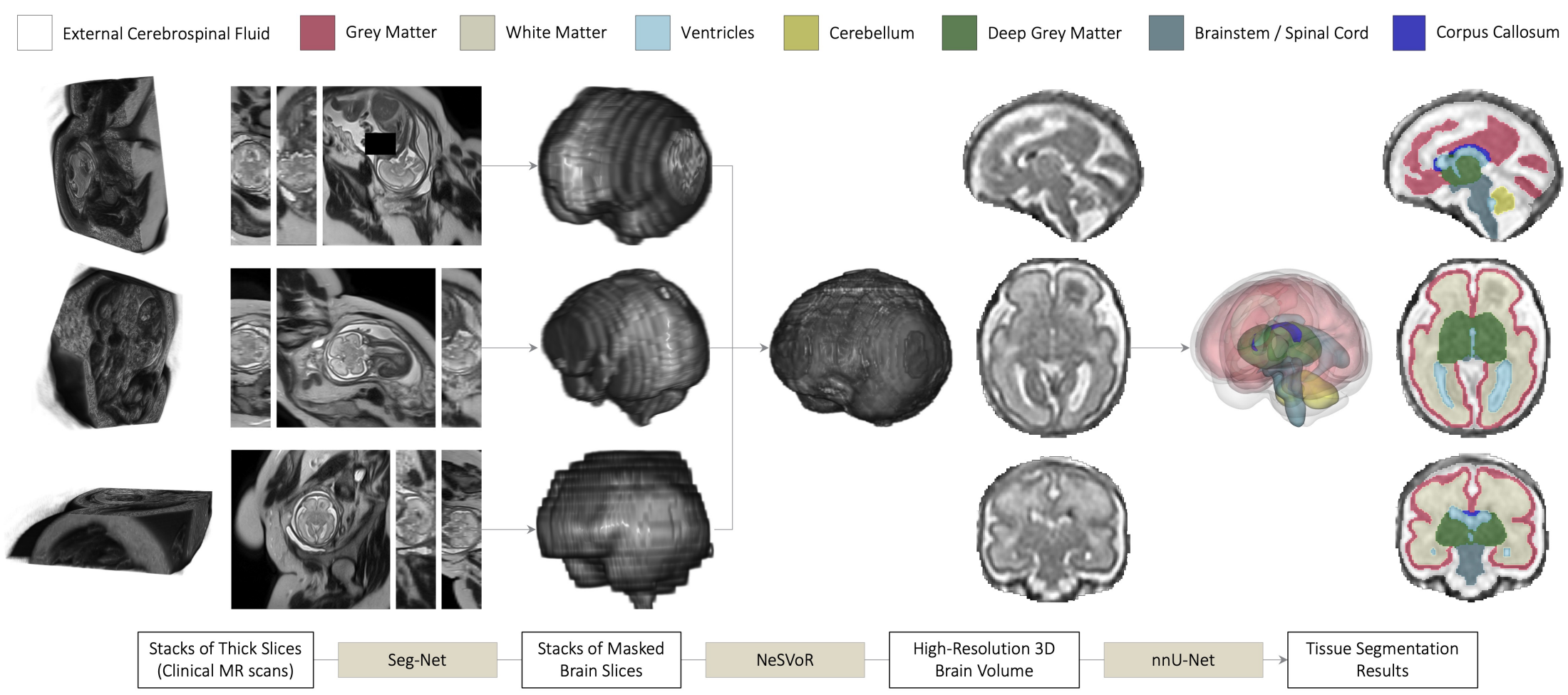

**Figure 2. Automated Pipeline for 3D Reconstruction and Tissue Segmentation of Fetal Brain MRI.** First, the fetal brain mask is extracted using the pre-trained neural network Seg-Net. Second, slice-to-volume reconstruction is performed with NeSVoR, based on implicit neural networks, to generate a high-resolution 3D brain volume. Last, tissue segmentation is performed using a pre-trained nnUNet.

### 2.2 Data Preprocessing and Annotation

Fetuses diagnosed with other brain abnormalities (e.g., agenesis of the corpus callosum, schizencephaly, cerebellar hypoplasia, and neuronal migration disorders) were excluded (Figure 1). All acquired imaging data were first converted from DICOM (Digital Imaging and Communications in Medicine) format to NifTI (Neuroimaging Informatics Technology Initiative) format using the DCM2NIIX software (V1.0.20220720). After that, data preprocessing consisted of two sequential steps: isotropic-resolution reconstruction and brain segmentation (Figure 2):

(1) Isotropic-resolution reconstruction. A pre-trained Seg-Net (Ebner et al., 2020) was used to localize the fetal brain in axial, sagittal, and coronal thick-slice image stacks for each subject. The GPU-accelerated NeSVoR (Xu et al., 2023) method was subsequently applied to reconstruct high-isotropic-resolution image volumes (0.8 × 0.8 × 0.8 mm³) from the three brain-masked thick-slice stacks. Briefly, NeSVoR performed motion correction by mapping 2D image slices into a 3D canonical space using the Slice-to-Volume Registration Transformer (SVoRT) (Xu et al., 2022), followed by volumetric reconstruction from multiple 2D slices via an implicit neural representation approach. The reconstructed volumes were zero-padded to a standardized matrix size of 210 × 210 × 210. Voxel intensities of each brain volume were normalized to the range [0, 1] using min–max normalization. All reconstructed 3D image volumes were visually reviewed for quality control by one investigator. Cases with unsuccessful volumetric reconstruction due to poor imaging quality (e.g., severe motion artifacts or incomplete anatomical coverage) were excluded (Figure 1).

(2) Brain segmentation. A trustworthy AI framework (Fidon et al., 2024) was employed to robustly segment each fetal brain volume into eight anatomical regions: external cerebrospinal fluid, gray matter, white matter, ventricles, cerebellum, deep gray matter, brainstem/spinal cord, and corpus callosum. This framework enhances the reliability of the AI model through a fallback strategy and a fail-safe mechanism based on Dempster–Shafer theory, enabling the pre-trained nnUNet segmentation model to generalize effectively across images acquired from different centers.

GMH-IVH was annotated and graded on reconstructed brain image volumes by two expert radiologists (E1 and E2 with 8 and 6 years of fetal imaging and diagnosis experience) to establish a reference standard. Hemorrhagic regions were first manually delineated by a radiologist (E2). The annotations were then reviewed and refined by the other radiologist (E1). In addition, the severity of GMH-IVH was evaluated and graded using a 1-4 grading scale (Papile et al., 1978) by a radiologist (E1), with Grade I for hemorrhage confined to the subependymal germinal matrix, Grade II for intraventricular hemorrhage, Grade III for hemorrhage with ventricular dilation or occupying more than 50% of the ventricle, and Grade IV for the presence of parenchymal hemorrhage.

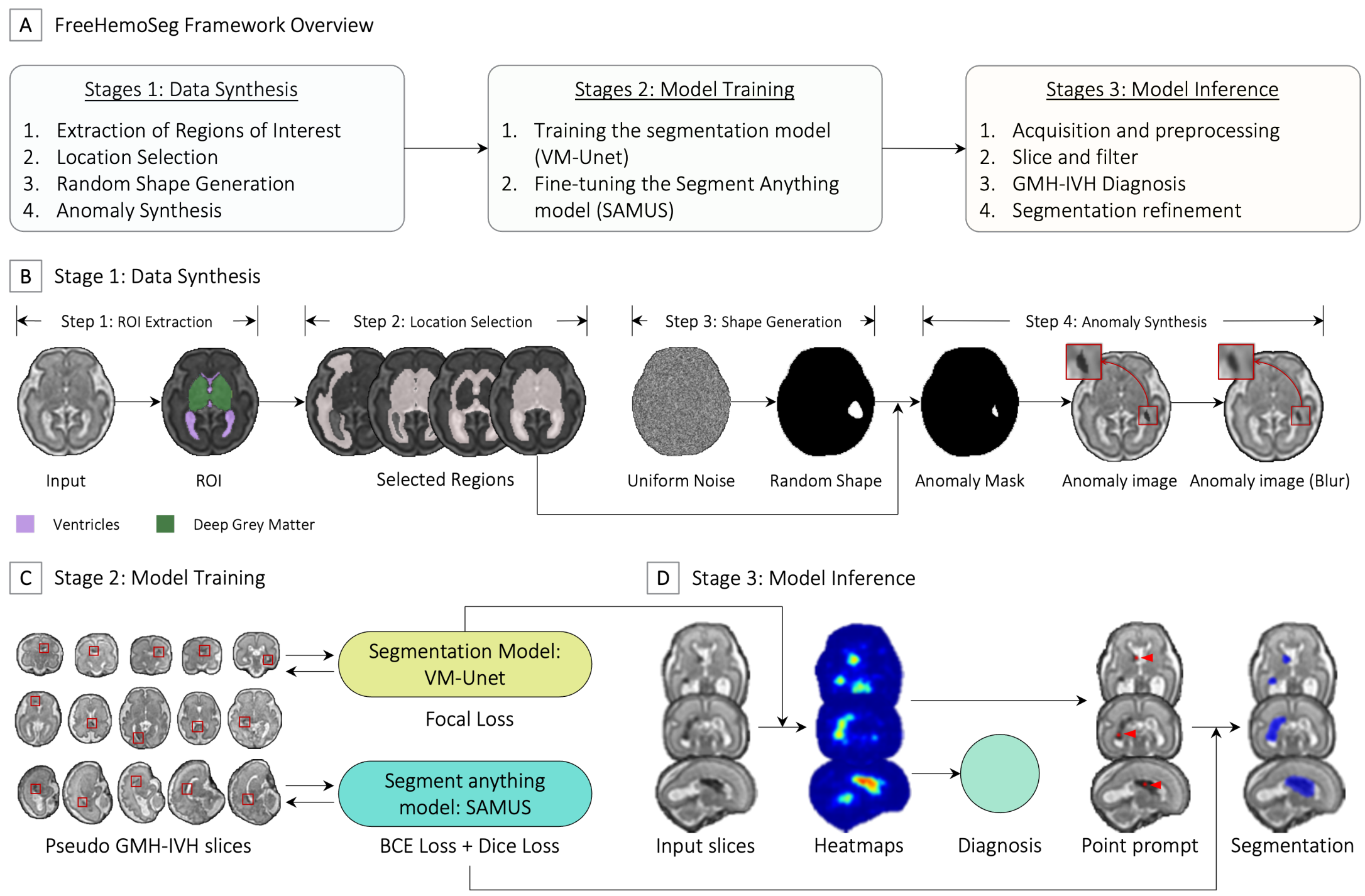

**Figure 3. FreeHemoSeg Framework.** (A) Overview of the three-stage pipeline. (B) Stage 1 – Data Synthesis: Pseudo GMH-IVH slices are synthesized from normal fetal brain images. (C) Stage 2 – Model Training: The synthesized slices are used to train segmentation models and fine-tune the SAM model. (D) Stage 3 – Model Inference: Segmentation probability heatmaps are generated for diagnosis and initial segmentation, and the coordinates of the highest-probability voxels in each slice are used as point prompts to guide the fine-tuned SAM model for further refinement.

## 2.3 Training and Validation Datasets

MRI data from one-fifth of randomly selected normal fetuses and two-thirds of randomly selected GMH-IVH cases from Hospital A were reserved for internal validation (Validation Dataset A), while the remaining data were used for training (Figure 1). Specifically, the GMH-IVH cases in the training data formed Training Dataset A, which was used to train a supervised learning model with labels. The normal cases in the training data comprised Training Dataset B, which was used to train UAD models. Using Training Dataset B as a base, the data synthesis process of FreeHemoSeg framework generated pseudo GMH-IVH data, which formed Training Dataset C for subsequent FreeHemoSeg model training. MRI data from Hospital B and Hospital C were used for external validation (Validation Dataset B). Notably, synthetic pseudo GMH-IVH images (Training Dataset C) were used solely for model training. All diagnostic and segmentation performance reported in this study was evaluated exclusively on real fetal MRI volumes from the internal and external validation datasets (Validation Datasets A and B).

## 2.4 FreeHemoSeg Framework Overview

FreeHemoSeg framework (Figure 3) aims to train deep learning models without manual annotations by constructing a training set through the synthesis of pseudo GMH-IVH slices from normal images. FreeHemoSeg consists of three stages (Figure 3A), including data synthesis, model training, and model inference:

(1) Data Synthesis. First, a large set of realistic synthetic pseudo GMH-IVH slices was generated from normal fetal brain images by applying Papile's grading system (Papile et al., 1978) and a random shape generation method that utilized uniform noise, Gaussian blurring, and morphological operations (Figure 3B, Section 2.5, Section 2.6).

(2) Model Training. Second, synthesized pseudo GMH-IVH slices were used to train a segmentation model, e.g., a VM-Unet (Ruan and Xiang, 2024) in this study, for coarse localization and to fine-tune a Segment Anything Model (SAM), a prompt-enabled foundation model pretrained on the SA-1B dataset (11+ million images, 1+ billion masks) with exceptional zero-shot generalization, for refined segmentation (Kirillov et al., 2023; Lin et al., 2024) (Figure 3C, online supplemental eFigure 1, eAppendix 1, and eAppendix 2).

(3) Model Inference. For inference, slices encompassing the ventricles and deep gray matter (DGM) of each fetus were processed by the trained VM-Unet, producing heatmaps from its final layer that represented hemorrhage probability for GMH-IVH diagnosis (Figure 3D). For case-level diagnosis, the maximum heatmap value across the entire brain volume was used as the anomaly score, and the fetus was classified as GMH-IVH if this score exceeded a predefined threshold. For slice-level diagnosis, the maximum heatmap value within each slice was used as the anomaly score, and the slice was classified as GMH-IVH if this score exceeded a predefined threshold. The coordinates of the maximum heatmap value within each slice were also used as point prompts to guide the fine-tuned SAM model, enabling refined and more accurate lesion segmentation.

### 2.5 Pseudo GMH-IVH Slices Synthesis

As illustrated in Figure 3B, the pseudo GMH-IVH slice synthesis pipeline comprises four steps:

(1) Slices were extracted from the reconstructed brain volumes along the axial, sagittal, and coronal planes. Based on tissue segmentation results, slices containing the ventricles, deep gray matter, and surrounding white matter were retained as regions of interest (ROIs). In addition, white matter alone was also considered as a potential alternative ROI (not shown in the figure).

(2) Potential hemorrhage regions were defined within the selected ROIs. According to established clinical knowledge, hemorrhages may occur in the ventricles and deep gray matter (Grades I–III) as well as in the surrounding white matter (Grade IV). Accordingly, four potential hemorrhage scenarios were designed: (i) The region of interest (ROI) was morphologically dilated to encompass both the ventricles and deep gray matter. (ii) The deep gray matter was excluded from the dilated ROI to simulate hemorrhage confined to the ventricles. (iii) The ventricular region was excluded to simulate hemorrhage localized to the deep gray matter. (iv) To model Grade IV hemorrhage, in which bleeding extends into the white matter, the white matter of either the left or right hemisphere was randomly selected. For each instance of pseudo-hemorrhage synthesis, these four scenarios were sampled with probabilities of 30%, 30%, 30%, and 10%, respectively.

(3) Random lesion shapes were generated using uniform noise followed by Gaussian blurring, spatial stretching, and thresholding, as described in Section 2.6.

(4) The random shape masks generated in step (3) were combined with the candidate hemorrhage regions defined in step (2) to form the final hemorrhage (anomaly) masks. Pixel intensities within each anomaly mask were then scaled by a factor ranging from 0.3 to 0.5 to generate initial abnormal images, reflecting the hypointense appearance of hemorrhage on T2-weighted MRI. Finally, Gaussian blurring was applied to further soften lesion boundaries, improving visual realism.

## 2.6 Random Shape Generation

Inspired by the approach of Zhang et al. (Zhang et al., 2023), a uniform-noise–based method was adopted for random shape generation (step 3 in Figure 3B). Given an input image $I \in R^{H\times W\times C}$ and a threshold $t \in R$, the objective is to generate a random shape mask $M \in \{0,1\}^{H\times W}$. First, a random noise image $N \in R^{H\times W}$ was generated with pixel intensities uniformly distributed between 0 and 255. The noise image was then smoothed using Gaussian blurring with a kernel size of 15 in both the horizontal and vertical directions, yielding a blurred image $B \in R^{H\times W}$. To maintain a consistent intensity range, $B$ was rescaled to [0, 255], producing an intermediate image $S \in R^{H\times W}$ with spatially varying intensity patterns. Next, a thresholding operation was applied to $S$ to obtain a binary image $T$. Morphological opening followed by closing was then performed on $T$ to remove small spurious regions and fill minor holes, resulting in a refined binary mask $R \in \{0,1\}^{H\times W}$. Finally, the output random shape mask $M$ was obtained by element-wise multiplication of $R$ with the brain mask $M_b \in \{0,1\}^{H\times W}$, ensuring that the generated shapes were confined within the brain region.

## 2.7 Comparison Methods

For GMH-IVH diagnosis, FreeHemoSeg was compared with three SOTA UAD methods trained on Training Dataset B, i.e., Skip-TS (Liu et al., 2024), RD4AD (Deng and Li, 2022), and IKD (Cao et al., 2022) (online supplemental eAppendix 3), as well as a VM-Unet (Ruan and Xiang, 2024) trained on Training Dataset A using supervised learning, i.e., SL Model. For GMH-IVH segmentation, FreeHemoSeg was compared against two methods: the SL Model and the FreeHemoSeg variant without SAM-based segmentation refinement (FreeHemoSeg w/o SAM).

**Table 2. Diagnostic Confidence Rating Scale for Fetal Germinal Matrix-intraventricular Hemorrhage.**

| Confidence Level | Quantitative Description |
|---|---|
| 1<br>(Very low) | Cannot reliably determine presence or absence of IVH; request supervised review. |
| 2<br>(Low) | May identify obvious GMH-IVH but remain uncertain for subtle or early bleeds; cannot confidently rule out GMH-IVH without confirmatory review. |
| 3<br>(Moderate) | Can detect moderate-to-large GMH-IVH and usually rule out clear negatives; may miss subtle/early GMH-IVH and seek second opinion. |
| 4<br>(High) | Reliably determine presence or absence of GMH-IVH. |
| 5<br>(Very high) | Confidently identify presence or absence of GMH-IVH, including subtle/early cases. |

### 2.8 Reader Study

To evaluate the clinical utility of FreeHemoSeg in assisting the diagnosis of fetal GMH–IVH, a reader study was conducted on Validation Dataset B (70 fetuses: 17 with GMH-IVH and 53 without). Two attending radiologists independently performed diagnostic assessments under three experimental conditions:

(1) 2D stack interpretation: Assessment based on raw 2D multiplanar image stacks (4–5 mm section thickness), reflecting the standard clinical workflow.

(2) 3D volume interpretation: Assessment based on 3D reconstructed brain volumes (0.8 mm isotropic) without AI assistance, evaluating the standalone benefit of volumetric reconstruction.

(3) FreeHemoSeg-assisted interpretation: Assessment based on 3D reconstructed brain volumes integrated with FreeHemoSeg-generated anomaly heatmaps.

Diagnostic performance was quantified using sensitivity and specificity. Diagnostic efficiency was evaluated by interpretation time, and diagnostic confidence was rated using a 5-point quantitative scale (Table 2).

## 2.9 Statistical Analysis

The performance was evaluated in terms of case-level diagnosis, slice-level diagnosis, and segmentation. For case-level diagnosis, the objective was to classify each fetus in the validation datasets as either GMH-IVH or Not GMH-IVH. For slice-level diagnosis, the goal was to classify each slice from all three anatomical planes (axial, coronal, and sagittal) of every fetus in the validation datasets as either GMH-IVH or Not GMH-IVH.

The diagnostic accuracy was evaluated using four metrics: the area under the receiver operating characteristic curve (AUROC), the area under the precision-recall curve (AUPR), sensitivity at a fixed specificity, and specificity at a fixed sensitivity. Sensitivity was evaluated at a high specificity of 0.8, aligning with the requirements of detection aid systems that prioritize minimizing false positives. Conversely, specificity was assessed at a high sensitivity of 0.8, reflecting the needs of triaging applications where the model functions as a prefilter (Lee et al., 2023). The segmentation performance was evaluated using dice similarity coefficients (DSCs) between the segmentation results and radiologist-annotated reference standard (RS).

Confidence intervals (CIs) for AUROC, AUPR and DSC were calculated via bootstrap (1000 resamples). CIs for sensitivity and specificity were determined using Wilson score intervals. Statistical comparisons employed the DeLong test for AUROC, bootstrap method for AUPR, paired Wilcoxon signed-rank test for DSC, and McNemar test for model performance at specific operating points. A *P* value less than .05 was considered to indicate a statistically significant difference. Data analysis was performed using Python version 3.8.13 (Python Software Foundation), and all metrics were calculated using the scikit-learn version 1.3 and NumPy version 1.24 packages.

**Table 3. Baseline Characteristics of the Study Population.** Abbreviations: UAD, unsupervised anomaly detection; GMH-IVH, germinal matrix-intraventricular hemorrhage; MA, maternal age; GA, gestational age; LP, left periventricular; RP, right periventricular; BP, bilateral periventricular.

| Characteristic | Training Dataset A | Training Dataset B | Validation Dataset A | Validation Dataset B |
|---|---|---|---|---|
| Function | SL Model training | UAD models training | Internal validation | External validation |
| No. of fetal brain volumes | | | | |
| GMH-IVH | 18 | 0 | 36 | 17 |
| Not GMH-IVH | 0 | 347 | 87 | 53 |
| No. of fetal brain slices | | | | |
| GMH-IVH | 653 | 0 | 1456 | 1410 |
| Not GMH-IVH | 0 | 40863 | 7177 | 6438 |
| GMH-IVH grade, n (%) | | | | |
| I | 12 (66.7%) | 0 | 25 (69.4%) | 1 (5.9%) |
| II | 4 (22.2%) | 0 | 6 (16.7%) | 3 (17.6%) |
| III | 1 (5.6%) | 0 | 3 (8.3%) | 6 (35.3%) |
| IV | 1 (5.6%) | 0 | 2 (5.6%) | 7 (41.2%) |
| MA, years, mean (SD), range | | | | |
| GMH-IVH | 26.6 (3.8), 17.0-33.0 | NA | 28.9 (4.0), 22.0-39.0 | 27.9 (2.6), 25.0-32.0 |
| Not GMH-IVH | NA | 28.3 (4.6), 19.0-41.0 | 28.8 (4.2), 20.0-43.0 | 30.9 (4.5), 19.0-40.0 |
| GA, weeks, mean (SD), range | | | | |
| GMH-IVH | 27.5 (3.9), 20.6-36.6 | NA | 28.6 (3.2), 23.6-34.3 | 30.0 (3.0), 24.4-35.6 |
| Not GMH-IVH | NA | 30.6 (3.5), 22.3-38.9 | 30.1 (3.4), 24.4-38.0 | 31.7 (3.4), 23.6-38.3 |
| Location of hemorrhagic lesions | | | | |
| LP | 6 (33.3%) | 0 | 13 (36.1%) | 5 (29.4%) |
| RP | 9 (50.0%) | 0 | 17 (47.2%) | 9 (52.9%) |
| BP | 3 (16.7%) | 0 | 6 (16.7%) | 3 (17.6%) |

## 3. Results

### 3.1 Data Characteristics

Among 1,797 stacks of 2D $T_2$-weighted images from 593 pregnant women across three hospitals, 123 stacks from 35 patients were excluded due to other brain abnormalities or poor image quality that prevented successful volumetric reconstruction. Therefore, the study dataset comprised 1,674 thick-slice stacks and the resulting 558 3D image volumes from 558 pregnant women (Figure 1). These volumes yielded a total of 57,997 axial, coronal, and sagittal 2D image slices covering the ventricles and DGM for analysis. The baseline demographic of the training and validation datasets are presented in Table 3. Notably, because the study was designed to detect GMH-IVH against a normal-brain background, fetuses with other structural brain abnormalities were excluded by design. The resulting low prevalence of brain anomalies in the final cohort therefore does not reflect the case mix of a general fetal MRI referral population.

Training Dataset A comprised 18 randomly selected GMH-IVH cases (653 brain slices), including 12 grade I (66.7%), 4 grade II (22.2%), and 1 each of grade III and grade IV (5.6%). Hemorrhage regions located in the left periventricular area in 6 cases (33.3%), the right in 9 cases (50.0%), and bilaterally in 3 cases (16.7%). Training Dataset C was derived from Training Dataset B and therefore shared identical demographic characteristics. Both datasets were composed of 40,863 brain slices from 347 brain volumes. The mean (SD) maternal age was 28.3 (4.6) years, and the mean (SD) gestational age was 30.6 (3.5) weeks (range, 22–39 weeks). Validation Dataset A comprised 87 normal and 36 GMH-IVH cases, while Validation Dataset B comprised 53 normal and 17 GMH-IVH cases.

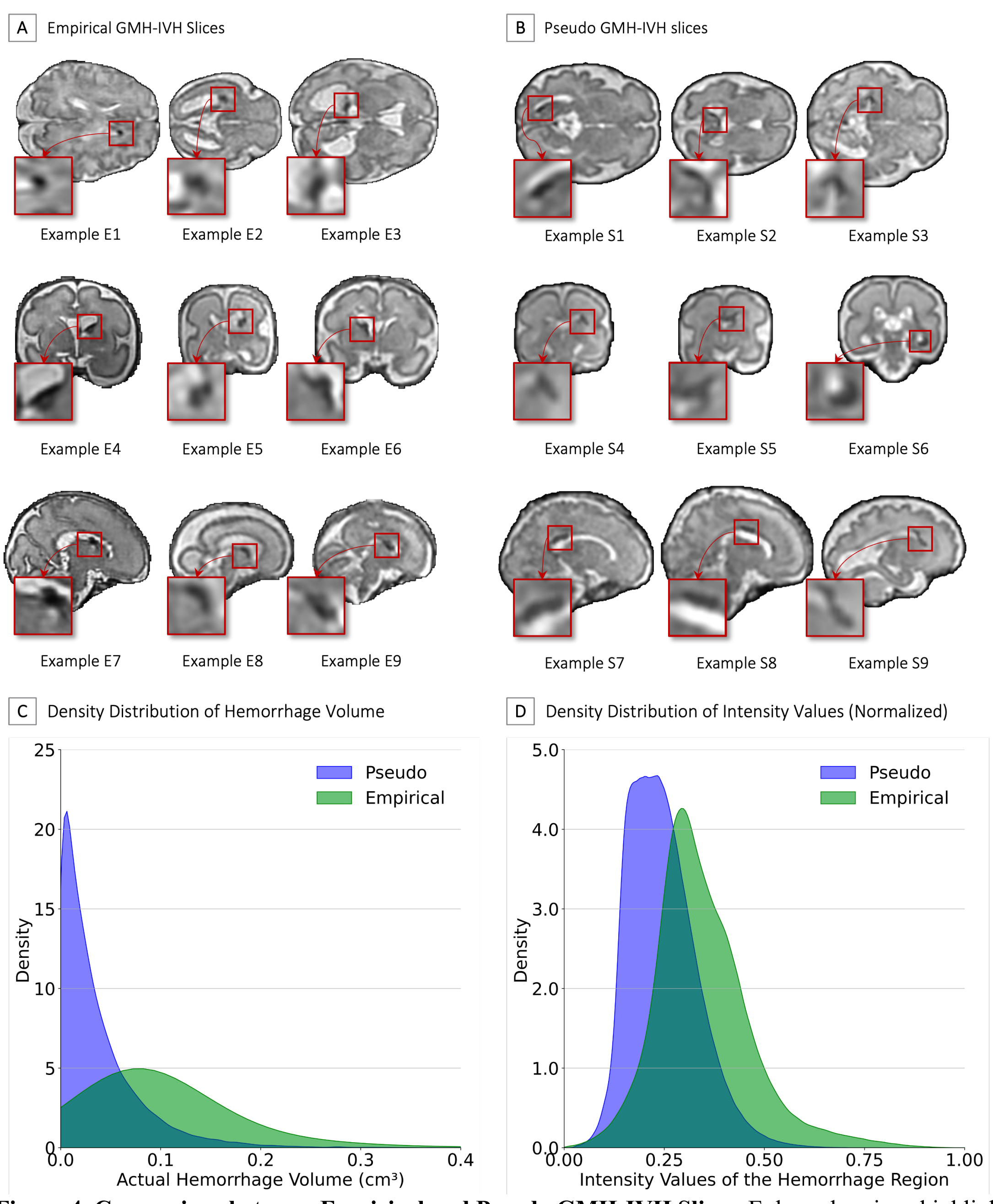

**Figure 4. Comparison between Empirical and Pseudo GMH-IVH Slices.** Enlarged regions highlight hemorrhagic areas of empirical GMH-IVH slices (A) and pseudo GMH-IVH slices (B). Slice-level comparison of hemorrhagic region sizes (C) and image intensity values (D) within the hemorrhage regions.

**Table 4. Diagnosis Accuracy.** Abbreviations: AUROC, area under the receiver operating characteristic curve; AUPR, area under the precision-recall curve; IKD, informative knowledge distillation; RD4AD, reverse distillation for anomaly detection; Skip-TS, teacher-student model with skip connections; SL Model, supervised learning model.

| **Internal Validation** | | | | | |
|---|---|---|---|---|---|
| **Method** | **Training Data** | **AUROC, (95% CI)** | **AUPR, (95% CI)** | **Sensitivity, (95% CI)** | **Specificity, (95% CI)** |
| Case-level Performance | | | | | |
| Skip-TS | Training Dataset B | 0.797 (0.694-0.890) | 0.699 (0.550-0.829) | 0.714 (0.650-0.771) | 0.747 (0.708-0.782) |
| RD4AD | Training Dataset B | 0.775 (0.675-0.864) | 0.569 (0.419-0.744) | 0.629 (0.561-0.691) | 0.586 (0.544-0.627) |
| IKD | Training Dataset B | 0.699 (0.595-0.795) | 0.543 (0.398-0.696) | 0.543 (0.475-0.609) | 0.437 (0.395-0.479) |
| SL Model | Training Dataset A | 0.829 (0.731-0.917) | 0.746 (0.598-0.871) | 0.743 (0.680-0.797) | 0.770 (0.732-0.804) |
| **FreeHemoSeg** | Training Dataset C | **0.959 (0.905-0.993)** | **0.928 (0.856-0.982)** | **0.914 (0.869-0.945)** | **0.966 (0.946-0.978)** |
| Slice-level Performance | | | | | |
| Skip-TS | Training Dataset B | 0.852 (0.840-0.863) | 0.608 (0.581-0.634) | 0.770 (0.761-0.779) | 0.763 (0.759-0.767) |
| RD4AD | Training Dataset B | 0.808 (0.797-0.819) | 0.428 (0.404-0.455) | 0.670 (0.660-0.679) | 0.683 (0.679-0.688) |
| IKD | Training Dataset B | 0.770 (0.757-0.783) | 0.422 (0.398-0.448) | 0.607 (0.597-0.617) | 0.575 (0.570-0.580) |
| SL Model | Training Dataset A | 0.785 (0.771-0.799) | 0.527 (0.500-0.553) | 0.648 (0.638-0.657) | 0.590 (0.585-0.594) |
| **FreeHemoSeg** | Training Dataset C | **0.910 (0.902-0.918)** | **0.760 (0.741-0.781)** | **0.844 (0.836-0.852)** | **0.852 (0.848-0.855)** |
| **External Validation** | | | | | |
| **Method** | **Training Data** | **AUROC, (95% CI)** | **AUPR, (95% CI)** | **Sensitivity, (95% CI)** | **Specificity, (95% CI)** |
| Case-level Performance | | | | | |
| Skip-TS | Training Dataset B | 0.776 (0.614-0.895) | 0.599 (0.372-0.813) | 0.588 (0.490-0.680) | 0.604 (0.549-0.656) |
| RD4AD | Training Dataset B | 0.605 (0.417-0.759) | 0.506 (0.285-0.720) | 0.353 (0.267-0.450) | 0.245 (0.202-0.295) |
| IKD | Training Dataset B | 0.675 (0.497-0.833) | 0.529 (0.303-0.743) | 0.529 (0.433-0.623) | 0.245 (0.201-0.296) |
| SL Model | Training Dataset A | 0.806 (0.703-0.899) | 0.536 (0.338-0.741) | 0.588 (0.491-0.679) | 0.660 (0.607-0.710) |
| **FreeHemoSeg** | Training Dataset C | **0.930 (0.837-0.993)** | **0.884 (0.751-0.982)** | **0.824 (0.739-0.885)** | **0.943 (0.913-0.964)** |
| Slice-level Performance | | | | | |
| Skip-TS | Training Dataset B | 0.861 (0.851-0.872) | 0.679 (0.657-0.702) | 0.762 (0.753-0.770) | 0.747 (0.742-0.751) |
| RD4AD | Training Dataset B | 0.788 (0.775-0.802) | 0.543 (0.517-0.569) | 0.632 (0.621-0.642) | 0.584 (0.579-0.588) |
| IKD | Training Dataset B | 0.785 (0.771-0.799) | 0.527 (0.500-0.557) | 0.630 (0.620-0.640) | 0.573 (0.568-0.578) |
| SL Model | Training Dataset A | 0.779 (0.766-0.794) | 0.536 (0.508-0.564) | 0.636 (0.626-0.646) | 0.552 (0.547-0.557) |
| **FreeHemoSeg** | Training Dataset C | **0.906 (0.896-0.915)** | **0.781 (0.763-0.799)** | **0.854 (0.847-0.861)** | **0.858 (0.855-0.862)** |

**Table 5. Significance of Diagnostic Accuracy Differences between FreeHemoSeg and Other Methods.** Abbreviations: AUROC, area under the receiver operating characteristic curve; AUPR, area under the precision-recall curve; IKD, informative knowledge distillation; RD4AD, reverse distillation for anomaly detection; Skip-TS, teacher-student model with skip connections; SL Model, supervised learning model.

| **Internal Validation** | | | | |
|---|---|---|---|---|
| **Method** | **AUROC** | **AUPR** | **Sensitivity** | **Specificity** |
| Case-level Performance | | | | |
| Skip-TS | 0.002 | 0.002 | 0.002 | 0.003 |
| RD4AD | 0.001 | < 0.001 | 0.001 | < 0.001 |
| IKD | < 0.001 | < 0.001 | < 0.001 | < 0.001 |
| SL Model | 0.002 | 0.011 | 0.003 | 0.004 |
| Slice-level Performance | | | | |
| Skip-TS | < 0.001 | < 0.001 | 0.082 | < 0.001 |
| RD4AD | < 0.001 | < 0.001 | < 0.001 | < 0.001 |
| IKD | < 0.001 | < 0.001 | < 0.001 | < 0.001 |
| SL Model | < 0.001 | < 0.001 | < 0.001 | < 0.001 |
| **External Validation** | | | | |
| **Method** | **AUROC** | **AUPR** | **Sensitivity** | **Specificity** |
| Case-level Performance | | | | |
| Skip-TS | 0.014 | 0.019 | 0.006 | < 0.001 |
| RD4AD | < 0.001 | 0.001 | 0.001 | < 0.001 |
| IKD | < 0.001 | 0.003 | 0.002 | < 0.001 |
| SL Model | 0.017 | 0.004 | 0.007 | < 0.001 |
| Slice-level Performance | | | | |
| Skip-TS | < 0.001 | < 0.001 | < 0.001 | < 0.001 |
| RD4AD | < 0.001 | < 0.001 | < 0.001 | < 0.001 |
| IKD | < 0.001 | < 0.001 | < 0.001 | < 0.001 |
| SL Model | < 0.001 | < 0.001 | < 0.001 | < 0.001 |

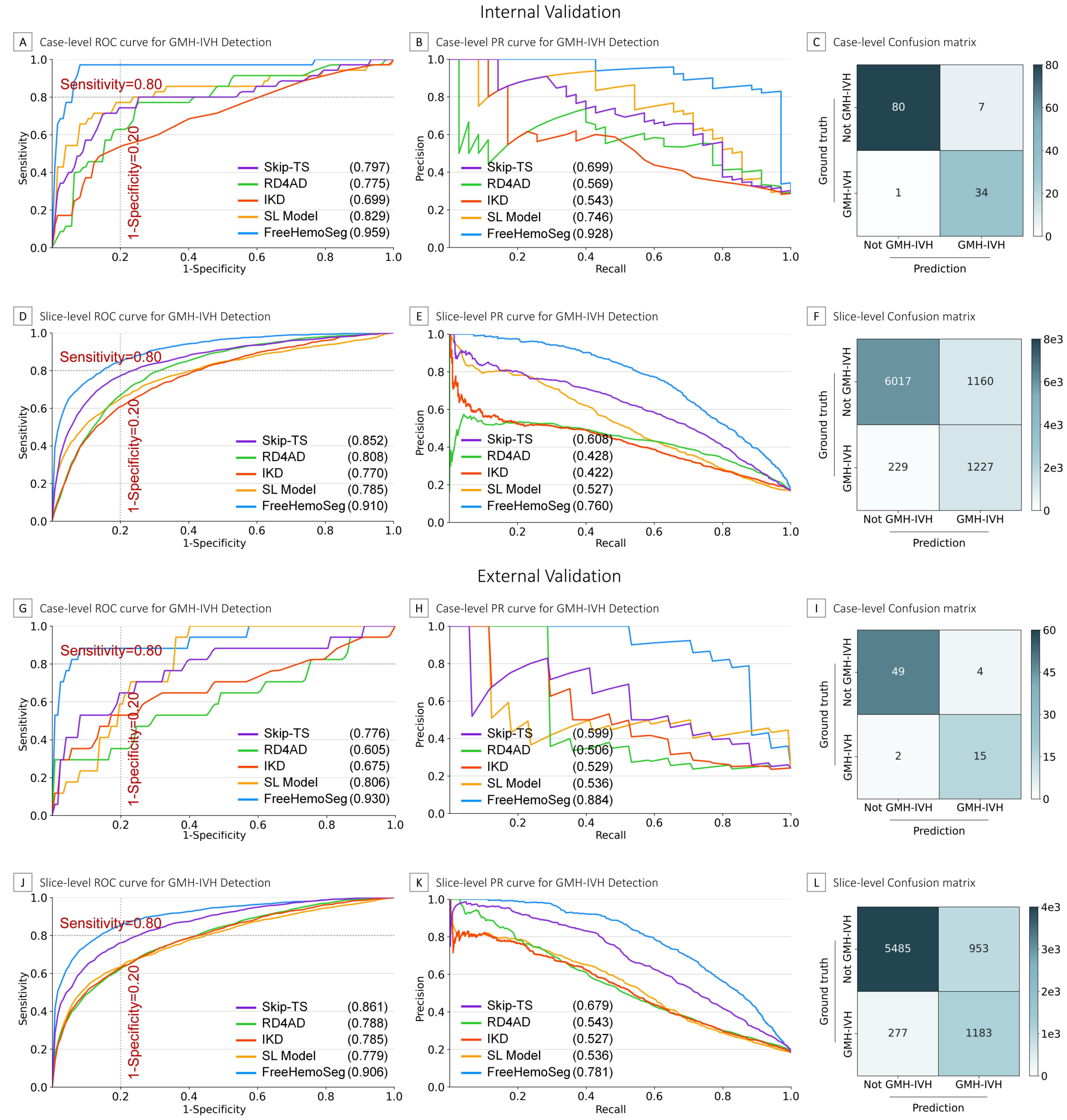

**Figure 5. Diagnosis Performance Analysis.** (A-F) Internal validation; (G-L) External validation. (A, G) Case-level receiver operating characteristic (ROC) curves for germinal matrix-intraventricular hemorrhage (GMH-IVH) diagnosis. (B, H) Case-level precision-recall (PR) curves for GMH-IVH diagnosis. (C, I) Confusion matrices showing case-level classification results for FreeHemoSeg at the threshold determined by maximizing Youden's index. (D, J) Slice-level ROC curves for GMH-IVH diagnosis. (E, K) Slice-level PR curves for GMH-IVH diagnosis. (F, L) Confusion matrices showing slice-level classification results for FreeHemoSeg at the threshold determined by Youden's index.

## 3.2 Model Performance

The synthesized GMH-IVH brain slices of FreeHemoSeg were consistent with MRI physics and anatomical priors, exhibiting characteristic low T2-weighted signal intensity and blurred lesion margins adjacent to the ventricles (Figure 4).

FreeHemoSeg achieved the best case-level and slice-level diagnostic accuracy on internal Validation Dataset A (Table 4, Table 5, Figure 5A-F). At the case level, it yielded an AUROC

of 0.959 (95% CI, 0.905–0.993), an AUPR of 0.928 (95% CI, 0.856–0.982), a sensitivity (at 0.8 specificity) of 0.914 (95% CI, 0.869–0.945), and a specificity (at 0.8 sensitivity) of 0.966 (95% CI, 0.946–0.978), significantly outperforming all other methods across these metrics (*P* < .05). At the slice level, FreeHemoSeg achieved an AUROC of 0.910 (95% CI, 0.902–0.918), an AUPR of 0.760 (95% CI, 0.741–0.781), a sensitivity of 0.844 (95% CI, 0.836–0.852), and a specificity of 0.852 (95% CI, 0.848–0.855), again outperforming all other methods (*P* < .05) except for the sensitivity of Skip-TS.

**Table 6. GMH-IVH Lesion Segmentation Performance.** Abbreviations: SL Model, supervised learning model; DSC, Dice similarity coefficient; SAM, segment anything model.

| **Internal Validation** | | | | | |
|---|---|---|---|---|---|
| **Method** | **Overall (*n* = 1456 slices)** | **Subgroup analysis** | | | |
| | | **Grade I (*n* = 914 slices)** | **Grade II (*n* = 271 slices)** | **Grade III (*n* = 123 slices)** | **Grade IV (*n* = 148 slices)** |
| DSC, (95% CI) | | | | | |
| SL Model | 0.479 (0.466, 0.493) | 0.482 (0.466, 0.498) | 0.463 (0.429, 0.497) | 0.237 (0.194, 0.280) | 0.693 (0.666, 0.721) |
| FreeHemoSeg w/o SAM | 0.526 (0.514, 0.538) | 0.518 (0.504, 0.532) | 0.476 (0.442, 0.510) | 0.543 (0.514, 0.572) | 0.649 (0.606, 0.692) |
| **FreeHemoSeg** | **0.559 (0.546, 0.571)** | **0.544 (0.530, 0.558)** | **0.520 (0.486, 0.553)** | **0.517 (0.472, 0.562)** | **0.754 (0.719, 0.789)** |
| *p*-value (vs. FreeHemoSeg) | | | | | |
| SL Model | < .001 | < .001 | < .001 | < .001 | < .001 |
| FreeHemoSeg w/o SAM | < .001 | < .001 | < .001 | 0.496 | < .001 |

| **External Validation** | | | | | |
|---|---|---|---|---|---|
| **Method** | **Overall (*n* = 1410 slices)** | **Subgroup analysis** | | | |
| | | **Grade I (*n* = 47 slices)** | **Grade II (*n* = 294 slices)** | **Grade III (*n* = 372 slices)** | **Grade IV (*n* = 697 slices)** |
| SL Model | 0.348 (0.335, 0.360) | 0.316 (0.279, 0.352) | 0.349 (0.327, 0.371) | 0.373 (0.347, 0.398) | 0.336 (0.317, 0.354) |
| FreeHemoSeg w/o SAM | 0.501 (0.487, 0.515) | **0.454 (0.392, 0.517)** | 0.563 (0.535, 0.590) | **0.531 (0.505, 0.557)** | 0.461 (0.440, 0.482) |
| **FreeHemoSeg** | **0.512 (0.497, 0.526)** | 0.430 (0.366, 0.495) | **0.574 (0.544, 0.605)** | 0.527 (0.501, 0.553) | **0.482 (0.460, 0.504)** |
| *p*-value (vs. FreeHemoSeg) | | | | | |
| SL Model | < .001 | < .001 | < .001 | < .001 | < .001 |
| FreeHemoSeg w/o SAM | 0.062 | 0.560 | 0.046 | 0.140 | 0.003 |

FreeHemoSeg also achieved the best case-level and slice-level diagnostic accuracy on external Validation Dataset B (Table 4, Table 5, Figure 5G-L). At the case level, it yielded an AUROC of 0.930 (95% CI, 0.837–0.993), an AUPR of 0.884 (95% CI, 0.751–0.982), a sensitivity of

0.824 (95% CI, 0.739–0.885), and a specificity of 0.943 (95% CI, 0.913–0.964), significantly outperforming all other methods across these metrics ($P < .05$). At the slice level, FreeHemoSeg achieved an AUROC of 0.906 (95% CI, 0.896–0.915), an AUPR of 0.781 (95% CI, 0.763–0.799), a sensitivity of 0.854 (95% CI, 0.847–0.861), and a specificity of 0.858 (95% CI, 0.855–0.862), again outperforming all other methods ($P < .05$).

Confusion matrices derived from the optimal classification thresholds, as determined by the Youden index, demonstrated that FreeHemoSeg accurately identified GMH-IVH at both slice and case levels (Figure 5C,F,I,L).

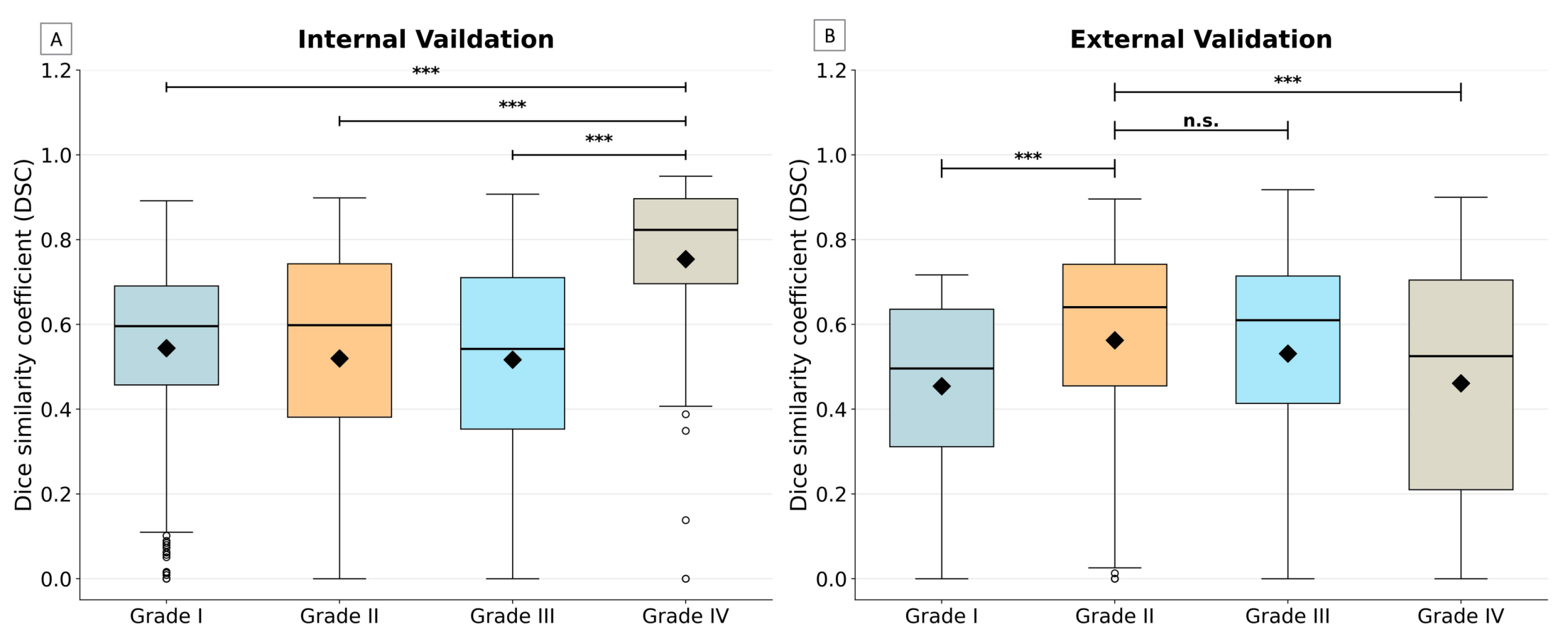

**Figure 6. Lesion Segmentation Performance of FreeHemoSeg across GMH-IVH Grades.** ***: $P < 0.001$

FreeHemoSeg achieved the best lesion segmentation performance (Table 6, Figure 6), achieving DSCs of 0.559 (95% CI, 0.546–0.571) and 0.512 (95% CI, 0.497–0.526) on the internal and external validation datasets, respectively. It significantly outperformed the SL Model in both datasets ($P < .05$). Compared with FreeHemoSeg without SAM, significant improvements were observed on the internal validation dataset ($P < .05$) and in Grade II and IV subgroups of the external validation dataset ($P < .05$).

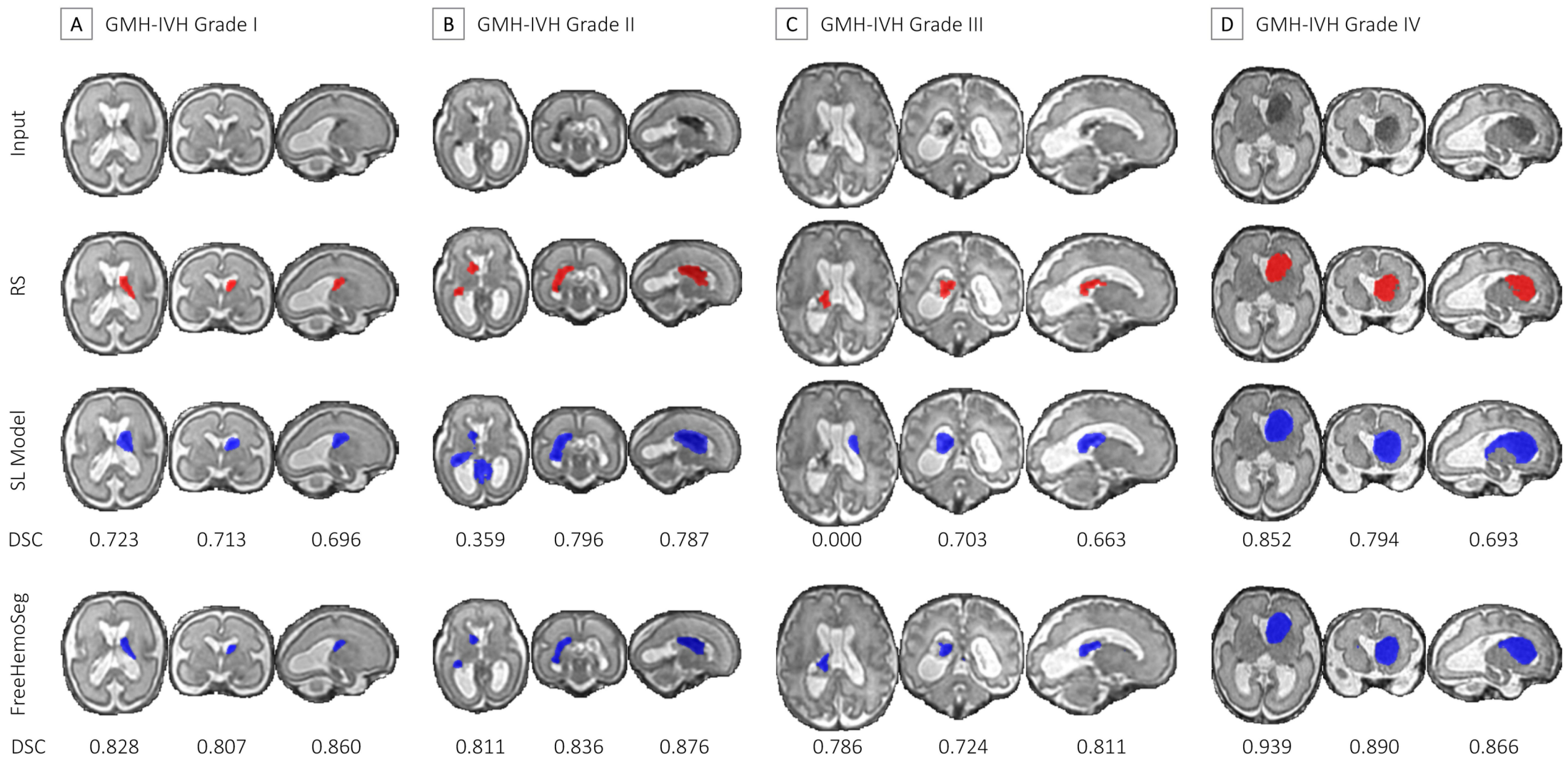

**Figure 7. Segmentation Visualization for Internal Validation.** $T_2$-weighted images, in which hemorrhagic lesions appear as low-intensity regions, along with lesion segmentation references and results from different methods, are shown for brains with: (A) Grade I hemorrhage confined to the germinal matrix, (B) Grade II hemorrhage extending into the ventricles, (C) Grade III hemorrhage penetrating the ventricles and causing marked ventricular dilation, and (D) Grade IV hemorrhage extending into the brain parenchyma. Abbreviations: RS, reference standard; SL Model, supervised learning model.

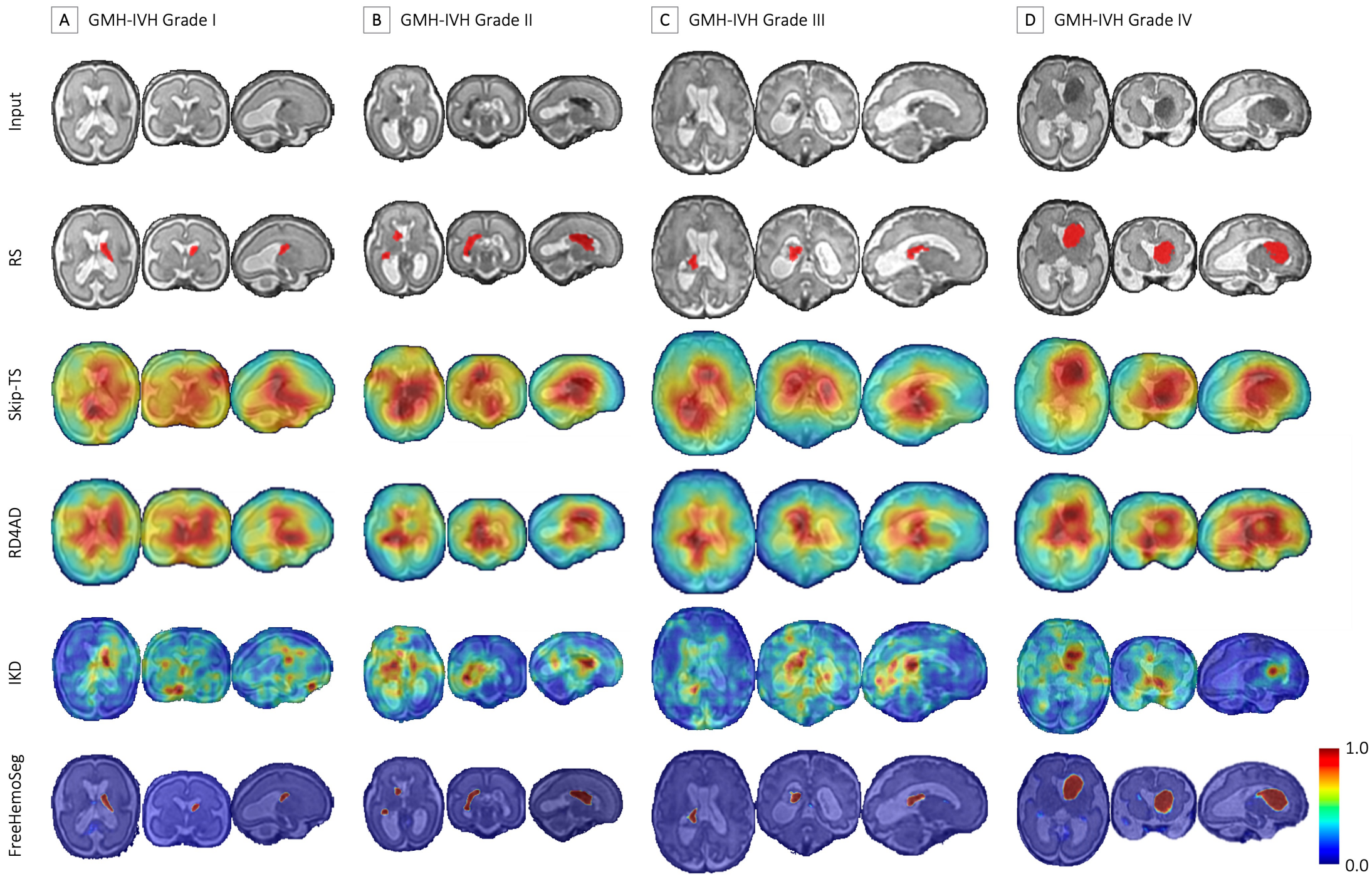

**Figure 8. Comparison of Output Heatmaps.** Comparison of heatmaps generated by FreeHemoSeg and other UAD models, including Skip-TS, RD4AD, IKD. Abbreviations: RS, reference standard; UAD, unsupervised anomaly detection; GMH-IVH, germinal matrix-intraventricular hemorrhage; IKD, informative knowledge distillation; RD4AD, reverse distillation for anomaly detection; Skip-TS, teacher-student model with skip connections.

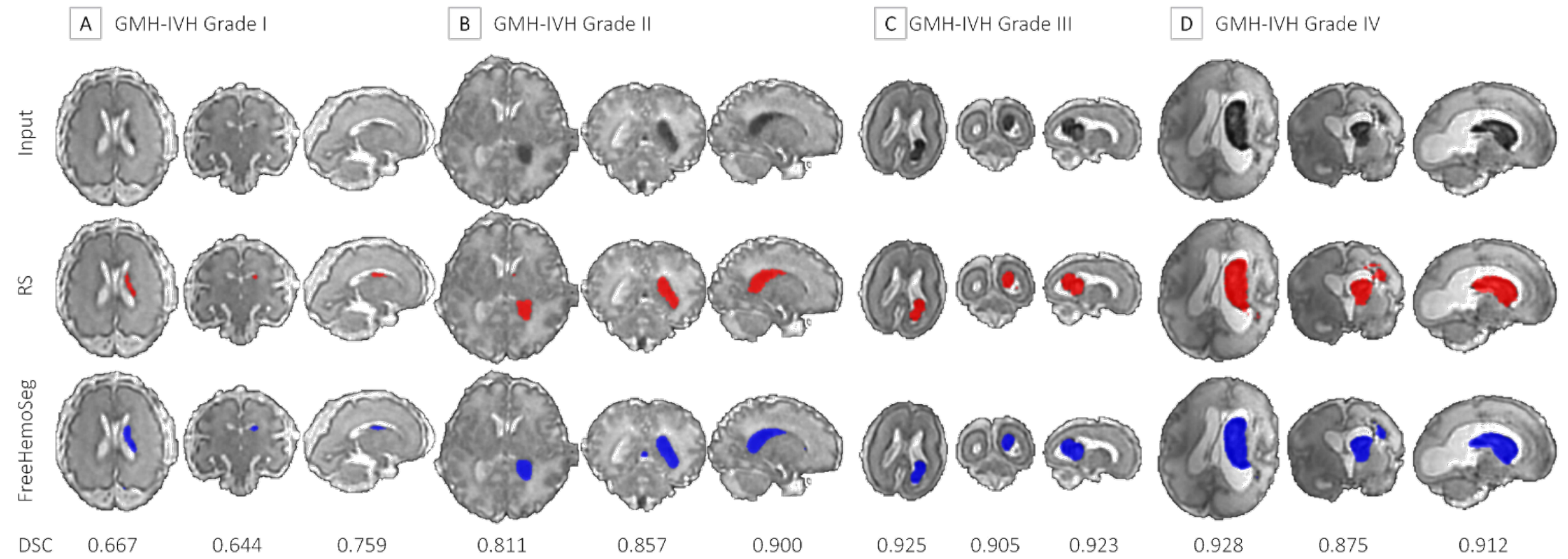

**Figure 9. Segmentation Visualization for External Validation.** $T_2$-weighted images, in which hemorrhagic lesions appear as low-intensity regions, along with lesion segmentation references and results from different methods, are shown for brains with: (A) Grade I hemorrhage confined to the germinal matrix, (B) Grade II hemorrhage extending into the ventricles, (C) Grade III hemorrhage penetrating the ventricles and causing marked ventricular dilation, and (D) Grade IV hemorrhage extending into the brain parenchyma. Abbreviations: RS, reference standard.

### 3.3 Segmentation Visualization

Visualization highlighted the differences in segmentation performance for cases with all four GMH-IVH grades across methods (Figure 7, Figure 8, and Figure 9). SL Model frequently over-segmented normal brain tissue and misidentified the germinal matrix as hemorrhage while missing actual hemorrhagic regions (axial plane, Figure 7C). Comparisons of heatmaps from the three SOTA annotation-free UAD models with FreeHemoSeg (Figure 8) further highlighted the limited ability of such methods to localize GMH-IVH lesions. Direct thresholding of these UAD heatmaps failed to produce plausible segmentations, so their DSC-based segmentation performance was not reported. Finally, visualizations on the Validation Dataset B demonstrated consistent segmentation performance of FreeHemoSeg on out-of-distribution images (Figure 9).

### 3.4 Segmentation Performance Improved by SAM in FreeHemoSeg

Owing to extensive pretraining on large-scale datasets and the use of point prompts to explicitly indicate target regions, SAM achieves superior segmentation performance after fine-tuning compared with end-to-end models such as Unet and VM-Unet. A practical limitation of SAM, however, is its reliance on manually specified point prompts. To address this issue, FreeHemoSeg automatically derived point prompts during the lesion segmentation stage by using the spatial coordinates corresponding to the maximum value in the anomaly heatmap generated by VM-Unet. These automatically generated prompts were then used to guide SAM, resulting in substantial performance gains.

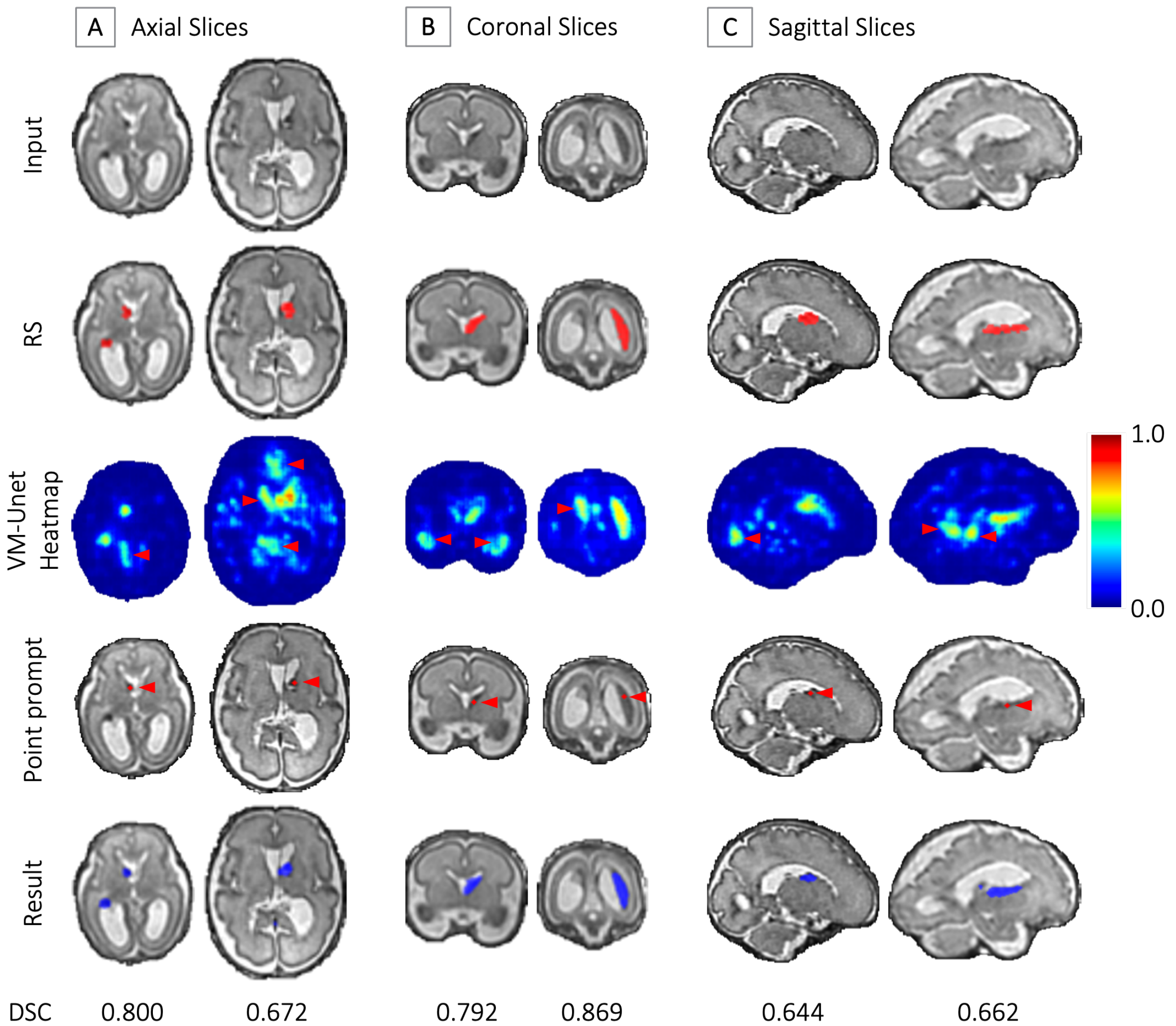

**Figure 10. Visualization of Enhanced False-positive Lesion Segmentation Results using Segment Anything Model (SAM).** First row: Input brain slices. Second row: Expert-annotated reference standard for hemorrhage regions. Third row: Heatmaps from VM-Unet showing high-probability areas for hemorrhage, with false positives in non-lesion tissues (red arrowheads). Fourth row: Point prompts generated from the coordinates of maximum values in the heatmap (red arrowheads). Fifth row: Fine-tuned SAM produced refined segmentation results based on the point prompts, correctly delineating hemorrhage regions. Abbreviations: RS, reference standard.

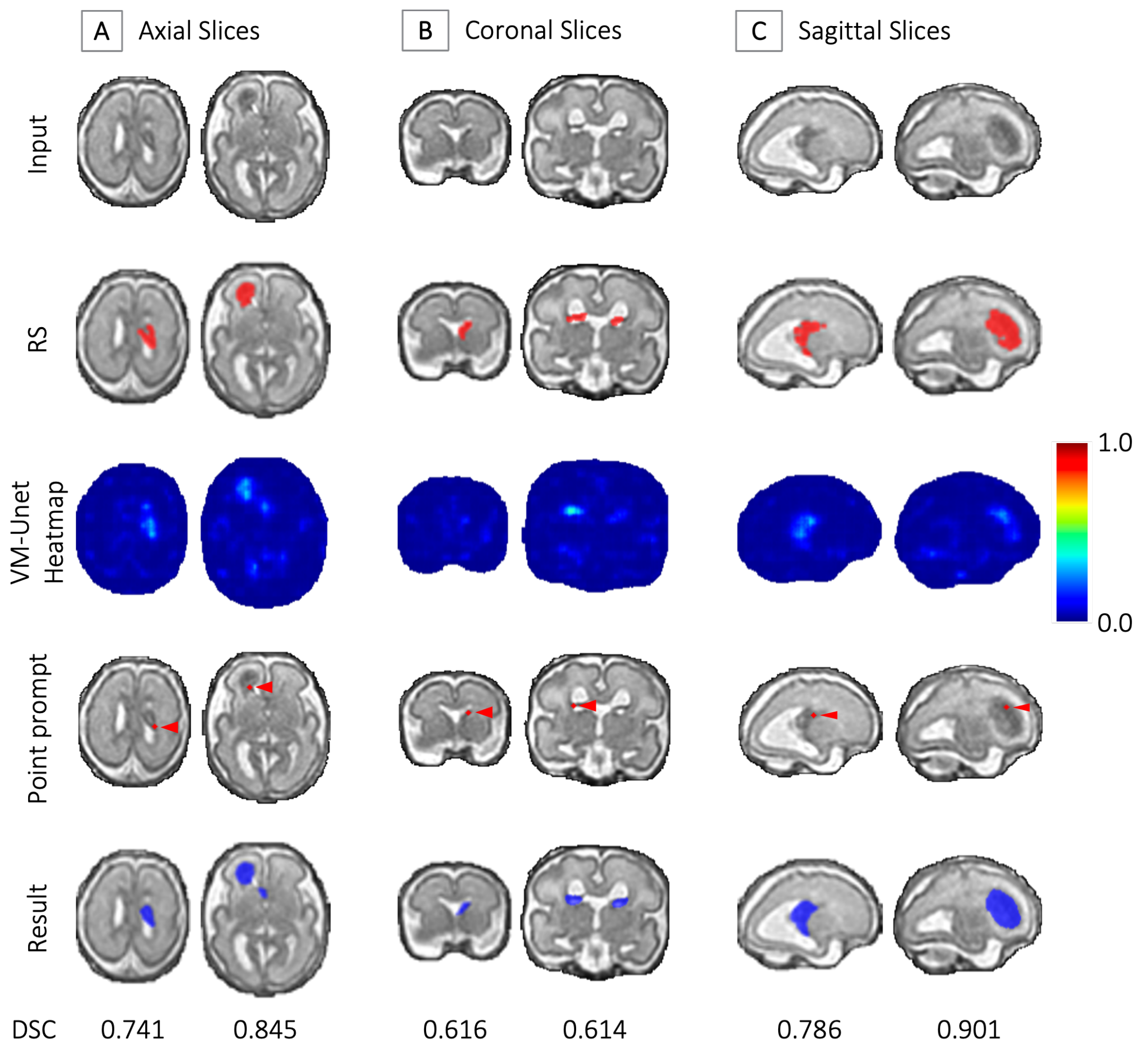

**Figure 11. Visualization of Enhanced False-negative Lesion Segmentation Results using Segment Anything Model (SAM).** First row: Input brain slices. Second row: Expert-annotated reference standard for hemorrhage regions. Third row: Heatmaps from VM-Unet showing high-probability areas for hemorrhage, with false negatives in some lesion regions. Fourth row: Point prompts generated from the coordinates of maximum values in the heatmap (red arrowheads). Fifth row: Fine-tuned SAM produced refined segmentation results based on the point prompts, correctly delineating hemorrhage regions. Abbreviations: RS, reference standard.

Visual examples illustrate how this strategy improved segmentation outcomes (Figure 10, Figure 11). First, VM-Unet-generated heatmaps may contain false-positive activations, with high-intensity responses in normal tissue regions (Figure 10). Direct thresholding of such heatmaps therefore led to reduced Dice similarity coefficient (DSC). In contrast, when a point prompt was generated at the location of the maximum heatmap value within the hemorrhagic region, SAM was able to disregard spurious activations in normal tissue and concentrate on the true lesion, thereby reducing false positives and producing a more accurate segmentation. Second, SAM also mitigated false-negative errors (Figure 11). In the illustrated examples, the VM-Unet heatmaps failed to exhibit high-intensity responses across the full extent of the hemorrhagic regions. Direct thresholding consequently resulted in under-segmentation or missed detections. However, provided that the automatically generated point prompt was located within the hemorrhagic region, SAM was able to recover the complete lesion extent and deliver accurate segmentation results.

### 3.5 Results of the Reader Study

FreeHemoSeg markedly improved diagnostic accuracy, efficiency, and confidence for two radiologists (Table 7). Compared with standard 2D stack interpretation, FreeHemoSeg's assistance increased sensitivity from 0.882 to 1.000 for R1 and from 0.882 to 0.941 for R2, while reducing interpretation time from 43.1 ± 28.1 s to 36.2 ± 17.4 s for R1 (16.0% reduction) and from 63.9 ± 21.9 s to 30.2 ± 7.5 s for R2 (52.7% reduction), with specificity remaining high across all conditions (≥ 0.943). Diagnostic confidence also increased for both readers (R1: from 4.74 to 4.83; R2: from 4.47 to 4.64), suggesting that the AI-generated heatmaps provided informative visual cues that facilitated reliable identification of both obvious and subtle hemorrhagic findings. Notably, 3D volume interpretation alone also reduced interpretation time and improved confidence relative to 2D stacks, but failed to improve and even decreased sensitivity (R1: 0.882; R2: 0.824), highlighting that the integration of FreeHemoSeg-generated heatmaps was critical for reducing missed diagnoses.

**Table 7. Diagnostic Performance with and without FreeHemoSeg Assistance on the External Validation Dataset**

| Radiologist | Method | Sensitivity | Specificity | Time (s) | Confidence |
|---|---|---|---|---|---|
| R1 | 2D Stacks | 0.882 | 1.000 | 43.1±28.1 | 4.74±0.50 |
| | 3D Volume | 0.882 | 1.000 | 22.8 ± 11.6 | 4.93 ± 0.26 |
| | FreeHemoSeg | 1.000 | 0.962 | 36.2±17.4 | 4.83±0.48 |
| R2 | 2D Stacks | 0.882 | 0.981 | 63.9±21.9 | 4.47±0.63 |
| | 3D Volume | 0.824 | 0.943 | 17.2 ± 1.9 | 4.79 ± 0.50 |
| | FreeHemoSeg | 0.941 | 0.962 | 30.2±7.5 | 4.64±0.56 |

## 4. Discussion

In this diagnostic study, we proposed the FreeHemoSeg framework, which generated GMH-IVH slices from normal fetal brain images using medical priors to train a deep learning algorithm in an annotation-free manner for GMH-IVH detection and hemorrhage region segmentation. The trained model was validated using radiologist-confirmed diagnostic results and segmentation masks as reference standards. FreeHemoSeg achieved the best performance on internal validation, with an AUROC of 0.959, an AUPR of 0.928, a sensitivity at fixed specificity of 0.914, and a specificity at fixed sensitivity of 0.966, as well as on external validation, with an AUROC of 0.930, an AUPR of 0.884, a sensitivity at fixed specificity of 0.824, and a specificity at fixed sensitivity of 0.943. FreeHemoSeg substantially improved the clinical diagnostic accuracy, efficiency, and confidence of radiologists in the reader study on the external Validation Dataset B.

The superior performance of the model trained on synthetic data can be attributed to the limited availability of prenatal GMH-IVH data for training, which restricts the comprehensiveness and diversity of empirical datasets. This scarcity hinders effective learning of both normal and hemorrhagic tissue characteristics. In contrast, the proposed GMH-IVH slice synthesis method can rapidly generate a diverse array of hemorrhage regions with varying shapes and distributions (Figure 4). To our knowledge, this study is the first to use an annotation-free deep learning framework for the simultaneous diagnosis and segmentation of fetal GMH-IVH. Automated assessment can be performed via a computer application (eAppendix 4, online supplemental eFigure 2 and eFigure 3), thereby streamlining the clinical workflow.

MRI is more sensitive than US for detecting prenatal GMH-IVH. US is a well-established method for prenatal diagnosis and typically depicts fetal GMH-IVH lesions as brightly echogenic areas with a homogeneous appearance (Sanapo et al., 2017). However, subtle and variable hemorrhage features on US make it challenging to distinguish fetal GMH-IVH from other intracranial lesions (Kim et al., 2024), with a low sensitivity of 60% even in postnatal cases (Alessandro Parodi and Ramenghi, 2015). By contrast, increasing evidence indicates that MRI allows more accurate identification and grading of hemorrhages missed by US (Dunbar et al., 2021; Elchalal et al., 2005; Kim et al., 2024; K.N. Epstein et al., 2021; Morioka et al., 2006; Ramenghi and Samuele, n.d.; Sanapo et al., 2017), which has been reported to change the counselling and/or management in approximately 22% of cases with central nervous system abnormalities (Gonçalves et al., 2016).

Nonetheless, high-resolution fetal MRI poses several challenges that must be addressed to improve the diagnostic accuracy of GMH-IVH. First, the large field of view and arbitrary fetal orientation result in inclusion of non-brain tissues and inconsistent brain positioning across slices (Salehi et al., 2018). Second, intermittent fetal motion produces thick-slice acquisitions, intra-slice artifacts, and inter-slice misalignment, hindering reliable volumetric analysis and interpretation. Last, the large data volume makes visual interpretation and lesion identification time-consuming and error-prone. FreeHemoSeg overcomes these limitations by performing automatic brain extraction with a deep learning-based fetal brain extractor, applying slice-to-volume reconstruction to mitigate motion artifacts (Xu et al., 2023), and delivering accurate diagnosis and lesion segmentation to assist clinicians. Additionally, while clinical guidelines (Prayer et al., 2023) recommend multi-sequence MRI to differentiate hemorrhages from other hypointense lesions, acquiring artifact-free $T_1$ and $T_2^*$ images in moving fetuses is technically demanding. FreeHemoSeg can achieve high diagnostic accuracy solely using standard $T_2$-weighted images, thereby simplifying the imaging protocol and reducing dependency on advanced sequence acquisition. We nonetheless recognize that susceptibility-sensitive sequences (e.g., echo planar imaging [EPI] and susceptibility-weighted imaging [SWI]) are intrinsically more sensitive to blood products. Extending the proposed annotation-free synthesis strategy to such sequences is an important next step.

Several considerations define where FreeHemoSeg is likely to add value. It is primarily intended to assist general radiologists and trainees with limited fetal MRI experience (the readers most likely to overlook subtle GMH-IVH), rather than to replace subspecialty fetal neuroradiologists. In routine practice it could act as a triage or second-reader tool that, after automatic brain extraction and slice-to-volume reconstruction, flags suspicious slices for review and provides segmentation and volume outputs to support structured reporting. Its incremental value is greatest when only $T_2$-weighted images are diagnostic. When full multi-sequence protocols are available, many mimics can already be differentiated confidently, and the tool's role shifts toward confirmation, quantification, and workflow efficiency. Finally, although developed for GMH-IVH, the annotation-free synthesis paradigm is potentially generalizable to other fetal brain pathologies with characteristic signal and anatomical priors, warranting future investigation.

This study has several limitations. (1) The synthesis of GMH-IVH slices incorporated only partial medical priors (hemorrhage location and T2-weighted hypointensity) and did not model hemorrhage size or the heterogeneous signal evolution that accompanies different blood ages, coagulation states, and surrounding tissue responses. As a result, the synthesized images still differed from real lesions (Figure 4C, 4D). We did not formally test whether expert fetal neuroradiologists could distinguish synthesized from authentic hemorrhages, and quantitative synthesis-fidelity metrics together with such reader-based assessments remain to be performed. (2) Although FreeHemoSeg achieved the best segmentation performance among the compared methods, the absolute DSCs (~0.51–0.56) indicate only moderate spatial overlap, and the segmentation output should therefore be regarded as a localization aid rather than a precise tool for volumetric quantification or grade determination. (3) The supervised comparator was trained on only 18 GMH-IVH cases (653 slices), reflecting the real-world scarcity of annotated fetal data that this work seeks to overcome. Its lower performance may partly reflect this scarcity rather than methodological inferiority, and the comparison should be interpreted accordingly. (4) The reference standard was established by expert radiological annotation rather than histopathological confirmation. Because microscopic blood degradation products may be present despite a normal imaging appearance, and because secondary blood products disseminating through the ventricular system or deposited along the ependymal lining may extend beyond the primary focus, the total hemorrhagic burden may be underestimated. (5) The dataset exhibited an imbalance in GMH-IVH severity grades, which may bias performance metrics. Larger prospective, multicentre studies are warranted, including high-risk referral populations, extension to susceptibility-sensitive sequences, and explicit analysis of pipeline robustness and error propagation.

## Conclusion

In conclusion, this study demonstrated that the proposed FreeHemoSeg framework achieves accurate performance in fetal GMH-IVH diagnosis and lesion segmentation without training on real patient data by integrating deep learning with medical prior-guided image synthesis. FreeHemoSeg addresses data scarcity and improves diagnostic accuracy, and may provide a foundation to support prognostic counselling and perinatal planning. Validation in larger prospective, multicentre cohorts, including high-risk populations and susceptibility-sensitive sequences, is needed before its potential roles in screening, grading, and prognostication can be established.

**Acknowledgements**

This work was supported by the National Natural Science Foundation of China (grant number 82572198), the Natural Science Foundation of Sichuan Province (grant number 2025ZNSFSC1768), the Scientific Research Project of Sichuan Medical Association (grant number 2024HR130), the Science and Technology Department of Sichuan Province, China (grant number 25SYSX0255), the Tsinghua University Startup Fund, and the Tsinghua University Dushi Program (grant numbers 20241080026, 20251080056).

**Data availability**

Raw fetal brain MRI data cannot be readily shared due to patient privacy constraints.

**Code availability**

The source codes of FreeHemoSeg are publicly available:

https://github.com/Arktis2022/FreeHemoSeg

The source codes of Skip-TS are publicly available:

https://github.com/Arktis2022/Skip-TS

The source codes of RD4AD are publicly available:

https://github.com/hq-deng/RD4AD

The source codes of IKD are publicly available:

https://github.com/caoyunkang/IKD

**References**

Alessandro Parodi, A.R., Giovanni Morana, Maria S. Severino, Mariya Malova, Anna R. Natalizia, Andrea Sannia, Ramenghi, L.A., 2015. Low-grade intraventricular hemorrhage: is ultrasound good enough? J. Matern. Fetal Neonatal Med. 28, 2261–2264. https://doi.org/10.3109/14767058.2013.796162

Bai, X., Liu, M., Chen, Y., Yang, H., Tian, Q., 2025. Chest-OMDL: Organ-specific Multidisease Detection and Localization in Chest Computed Tomography using Weakly Supervised Deep Learning from Free-text Radiology Report, in: Medical Imaging with Deep Learning.

Brouwer, A.J., Groenendaal, F., Benders, M.J.N.L., De Vries, L.S., 2014. Early and Late Complications of Germinal Matrix-Intraventricular Haemorrhage in the Preterm Infant: What Is New? Neonatology 106, 296–303. https://doi.org/10.1159/000365127

Cao, Y., Wan, Q., Shen, W., Gao, L., 2022. Informative knowledge distillation for image anomaly segmentation. Knowl.-Based Syst. 248, 108846. https://doi.org/10.1016/j.knosys.2022.108846

Chalcroft, L., Pappas, I., Price, C.J., Ashburner, J., 2024. Synthetic Data for Robust Stroke Segmentation.

Chen, Y., 2017. A Labeling-Free Approach to Supervising Deep Neural Networks for Retinal Blood Vessel Segmentation.

Deng, H., Li, X., 2022. Anomaly Detection via Reverse Distillation from One-Class Embedding, in: 2022 IEEE/CVF Conference on Computer Vision and Pattern Recognition (CVPR). Presented at the 2022 IEEE/CVF Conference on Computer Vision and Pattern Recognition (CVPR), IEEE, New Orleans, LA, USA, pp. 9727–9736. https://doi.org/10.1109/CVPR52688.2022.00951

Dunbar, M.J., Woodward, K., Leijser, L.M., Kirton, A., 2021. Antenatal diagnosis of fetal intraventricular hemorrhage: systematic review and meta-analysis. Dev. Med. Child Neurol. 63, 144–155. https://doi.org/10.1111/dmcn.14713

Ebner, M., Wang, G., Li, W., Aertsen, M., Patel, P.A., Aughwane, R., Melbourne, A., Doel, T., Dymarkowski, S., De Coppi, P., David, A.L., Deprest, J., Ourselin, S., Vercauteren, T., 2020. An automated framework for localization, segmentation and super-resolution reconstruction of fetal brain MRI. NeuroImage 206, 116324. https://doi.org/10.1016/j.neuroimage.2019.116324

Elchalal, U., Yagel, S., Gomori, J.M., Porat, S., Beni-Adani, L., Yanai, N., Nadjari, M., 2005. Fetal intracranial hemorrhage (fetal stroke): does grade matter? Ultrasound Obstet. Gynecol. 26, 233–243. https://doi.org/10.1002/uog.1969

Fidon, L., Aertsen, M., Kofler, F., Bink, A., David, A.L., Deprest, T., Emam, D., Guffens, F., Jakab, A., Kasprian, G., Kienast, P., Melbourne, A., Menze, B., Mufti, N., Pogledic, I., Prayer, D., Stuempflen, M., Van Elslander, E., Ourselin, S., Deprest, J., Vercauteren, T., 2024. A Dempster-Shafer approach to trustworthy AI with application to fetal brain MRI segmentation. IEEE Trans. Pattern Anal. Mach. Intell. 1–12. https://doi.org/10.1109/TPAMI.2023.3346330

Gonçalves, L.F., Lee, W., Mody, S., Shetty, A., Sangi-Haghpeykar, H., Romero, R., 2016. Diagnostic accuracy of ultrasonography and magnetic resonance imaging for the detection of fetal anomalies: a blinded case-control study: Accuracy of US and MRI in detecting fetal anomalies. Ultrasound Obstet. Gynecol. 48, 185–192. https://doi.org/10.1002/uog.15774

Hadi, E., Haddad, L., Levy, M., Gindes, L., Hausman-Kedem, M., Bassan, H., Ben-Sira, L., Libzon, S., Kassif, E., Hoffmann, C., Leibovitz, Z., Kasprian, G., Lerman-Sagie, T., 2024. Fetal intraventricular hemorrhage and periventricular hemorrhagic venous infarction: time for dedicated classification system. Ultrasound Obstet. Gynecol. Off. J. Int. Soc. Ultrasound Obstet. Gynecol. 64, 285–293. https://doi.org/10.1002/uog.27613

Hao, Y., Liu, M., Zhu, J., Yang, H., Li, H., Kang, M., Song, Y., Lai, H., Zhou, X., Ning, G., Liao, Y., Qu, H., Tian, Q., 2026. PANDA: Patch-based Unsupervised Deep Learning for Brain Anomaly Detection via Age Prediction in Fetal MRI. Imaging Neurosci. https://doi.org/10.1162/IMAG.a.1293

Huang, H., Lin, L., Tong, R., Hu, H., Zhang, Q., Iwamoto, Y., Han, X., Chen, Y.-W., Wu, J., 2020. UNet 3+: A Full-Scale Connected UNet for Medical Image Segmentation, in: ICASSP 2020 - 2020 IEEE International Conference on Acoustics, Speech and Signal Processing (ICASSP). Presented at the ICASSP 2020 - 2020 IEEE International Conference on Acoustics, Speech and Signal Processing (ICASSP), IEEE, Barcelona, Spain, pp. 1055–1059. https://doi.org/10.1109/ICASSP40776.2020.9053405

Kim, S., Jung, Y.J., Baik, J., Kwon, H., Lee, J., Kwon, J.-Y., Kim, Y.-H., 2024. Prenatal diagnosis and postnatal outcome of fetal intracranial hemorrhage: a single-center experience. Obstet. Gynecol. Sci. 67, 393–403. https://doi.org/10.5468/ogs.24097

Kirillov, A., Mintun, E., Ravi, N., Mao, H., Rolland, C., Gustafson, L., Xiao, T., Whitehead, S., Berg, A.C., Lo, W.-Y., 2023. Segment anything, in: Proceedings of the IEEE/CVF International Conference on Computer Vision. pp. 4015–4026.

K.N. Epstein, Kline-Fath, B.M., Zhang, B., Venkatesan, C., Habli, M., Dowd, D., Nagaraj, U.D., 2021. Prenatal Evaluation of Intracranial Hemorrhage on Fetal MRI: A Retrospective Review. Am. J. Neuroradiol. 42, 2222–2228. https://doi.org/10.3174/ajnr.A7320

Lee, W., Lee, Hyeonsoo, Lee, Hyunjae, Park, E.K., Nam, H., Kooi, T., 2023. Transformer-based Deep Neural Network for Breast Cancer Classification on Digital Breast Tomosynthesis Images. Radiol. Artif. Intell. 5, e220159. https://doi.org/10.1148/ryai.220159

Li, X., Chen, H., Qi, X., Dou, Q., Fu, C.-W., Heng, P.-A., 2018. H-DenseUNet: Hybrid Densely Connected UNet for Liver and Tumor Segmentation From CT Volumes. IEEE Trans. Med. Imaging 37, 2663–2674. https://doi.org/10.1109/TMI.2018.2845918

Lin, X., Xiang, Y., Yu, L., Yan, Z., 2024. Beyond Adapting SAM: Towards End-to-End Ultrasound Image Segmentation via Auto Prompting, in: Linguraru, M.G., Dou, Q., Feragen, A., Giannarou, S., Glocker, B., Lekadir, K., Schnabel, J.A. (Eds.), Medical

Image Computing and Computer Assisted Intervention – MICCAI 2024. Springer Nature Switzerland, Cham, pp. 24–34.

Liu, M., Jiao, Y., Lu, J., Chen, H., 2024. Anomaly Detection for Medical Images Using Teacher-Student Model with Skip Connections and Multi-scale Anomaly Consistency. IEEE Trans. Instrum. Meas. 1–1. https://doi.org/10.1109/TIM.2024.3406792

Liu, M., Liao, Y., Li, H., Zhu, J., Yang, H., Hao, Y., Qu, H., Tian, Q., 2026. Quality-label-free Fetal Brain MRI Quality Control Based on Image Orientation Recognition Uncertainty. Med. Image Anal. 103994. https://doi.org/https://doi.org/10.1016/j.media.2026.103994

Lyu, F., Ye, M., Yip, T.C.-F., Wong, G.L.-H., Yuen, P.C., 2024. Local Style Transfer via Latent Space Manipulation for Cross-Disease Lesion Segmentation. IEEE J. Biomed. Health Inform. 28, 273–284. https://doi.org/10.1109/JBHI.2023.3327726

Moradi, B., Ardestani, R.M., Shirazi, M., Eslamian, L., Kazemi, M.A., 2024. Fetal intracranial hemorrhage and infarct: Main sonographic and MRI characteristics: A review article. Eur. J. Obstet. Gynecol. Reprod. Biol. X 24, 100351. https://doi.org/10.1016/j.eurox.2024.100351

Morioka, T., Hashiguchi, K., Nagata, S., Miyagi, Y., Mihara, F., Hikino, S., Tsukimori, K., Sasaki, T., 2006. Fetal Germinal Matrix and Intraventricular Hemorrhage. Pediatr. Neurosurg. 42, 354–361. https://doi.org/10.1159/000095565

Papile, L.-A., Burstein, J., Burstein, R., Koffler, H., 1978. Incidence and evolution of subependymal and intraventricular hemorrhage: a study of infants with birth weights less than 1,500 gm. J. Pediatr. 92, 529–534.

Prayer, D., Malinger, G., De Catte, L., De Keersmaecker, B., Gonçalves, L.F., Kasprian, G., Laifer-Narin, S., Lee, W., Millischer, A. -E., Platt, L., Prayer, F., Pugash, D., Salomon, L.J., Sanz Cortes, M., Stuhr, F., Timor-Tritsch, I.E., Tutschek, B., Twickler, D., Raine-Fenning, N., the ISUOG Clinical Standards Committee, 2023. ISUOG Practice Guidelines (updated): performance of fetal magnetic resonance imaging. Ultrasound Obstet. Gynecol. 61, 278–287. https://doi.org/10.1002/uog.26129

Ramenghi, L.A., Samuele, C., n.d. FETAL INTRAVENTRICULAR HAEMORRHAGE: DIFFERENT APPROACHES NEEDED? INSIGHTS ON GASLINI'S EXPERIENCE.

Ruan, J., Xiang, S., 2024. VM-UNet: Vision Mamba UNet for Medical Image Segmentation.

Salehi, S.S.M., Hashemi, S.R., Velasco-Annis, C., Ouaalam, A., Estroff, J.A., Erdogmus, D., Warfield, S.K., Gholipour, A., 2018. Real-time automatic fetal brain extraction in fetal MRI by deep learning, in: 2018 IEEE 15th International Symposium on Biomedical Imaging (ISBI 2018). pp. 720–724. https://doi.org/10.1109/ISBI.2018.8363675

Sanapo, L., Whitehead, M.T., Bulas, D.I., Ahmadzia, H.K., Pesacreta, L., Chang, T., Du Plessis, A., 2017. Fetal intracranial hemorrhage: role of fetal MRI: Fetal intracranial hemorrhage and fetal MRI. Prenat. Diagn. 37, 827–836. https://doi.org/10.1002/pd.5096

Wang, Y., Chen, Y., Jiang, S., Yu, W., Liu, M., Wu, B., Zhu, S., Qin, F., Fan, J., Wang, C., 2025a. DR-TTA: Dynamic and Robust Test-Time Adaptation Under Low-Quality Mri Conditions for Brain Tumor Segmentation, in: 2025 IEEE International Conference on Bioinformatics and Biomedicine (BIBM). pp. 2899–2906. https://doi.org/10.1109/BIBM66473.2025.11356381

Wang, Y., Chen, Y., Jiang, S., Yu, W., Liu, M., Wu, B., Zong, J., Qin, F., Wang, C., Tian, Q., 2025b. SmaRT: Style-Modulated Robust Test-Time Adaptation for Cross-Domain Brain Tumor Segmentation in MRI. ArXiv E-Prints arXiv:2509.17925. https://doi.org/10.48550/arXiv.2509.17925

Xu, J., Moyer, D., Gagoski, B., Iglesias, J.E., Grant, P.E., Golland, P., Adalsteinsson, E., 2023. NeSVoR: Implicit Neural Representation for Slice-to-Volume Reconstruction in MRI. IEEE Trans. Med. Imaging 42, 1707–1719. https://doi.org/10.1109/TMI.2023.3236216

Xu, J., Moyer, D., Grant, P.E., Golland, P., Iglesias, J.E., Adalsteinsson, E., 2022. SVoRT: Iterative Transformer for Slice-to-Volume Registration in Fetal Brain MRI, in: Wang, L., Dou, Q., Fletcher, P.T., Speidel, S., Li, S. (Eds.), Medical Image Computing and Computer Assisted Intervention – MICCAI 2022, Lecture Notes in Computer Science.

Springer Nature Switzerland, Cham, pp. 3–13. https://doi.org/10.1007/978-3-031-16446-0_1

Yao, Q., Xiao, L., Liu, P., Zhou, S.K., 2021. Label-Free Segmentation of COVID-19 Lesions in Lung CT. IEEE Trans. Med. Imaging 40, 2808–2819. https://doi.org/10.1109/TMI.2021.3066161

Zhang, H., Yang, J., Wan, S., Fua, P., 2024. LeFusion: Synthesizing Myocardial Pathology on Cardiac MRI via Lesion-Focus Diffusion Models.

Zhang, Z., Deng, H., Li, X., 2023. Unsupervised Liver Tumor Segmentation with Pseudo Anomaly Synthesis, in: Wolterink, J.M., Svoboda, D., Zhao, C., Fernandez, V. (Eds.), Simulation and Synthesis in Medical Imaging, Lecture Notes in Computer Science. Springer Nature Switzerland, Cham, pp. 86–96. https://doi.org/10.1007/978-3-031-44689-4_9

Zhou, K., Li, J., Luo, W., Li, Z., Yang, J., Fu, H., Cheng, J., Liu, J., Gao, S., 2022. Proxy-Bridged Image Reconstruction Network for Anomaly Detection in Medical Images. IEEE Trans. Med. Imaging 41, 582–594. https://doi.org/10.1109/TMI.2021.3118223

Zong, J., Chen, Y., Liu, M., Du, Y., Wu, C., Wu, B., Zhou, G., Qin, F., 2026. MUIT-TTA: Annotation-free intracranial hemorrhage segmentation via pseudo-anomaly synthesis and test-time adaptation. Pattern Recognit. 114143. https://doi.org/https://doi.org/10.1016/j.patcog.2026.114143

**Supplementary information**

**eAppendix 1. Detailed VM-Unet Architecture and Training Procedures**

A VM-Unet model (Ruan and Xiang, 2024) was trained for coarse localization of GMH-IVH, taking single-channel images of size 256 × 256 × 1 as input. VM-Unet is a variant of the Unet architecture and retains the characteristic encoder–decoder structure with downsampling and upsampling paths, while incorporating an enhanced backbone network (eFigure 1). Specifically, the architecture begins with a patch embedding layer that partitions the input image into non-overlapping 4 × 4 patches. The encoder then performs hierarchical feature extraction using stacked Visual State Space (VSS) blocks derived from VMamba (Y. Liu et al., 2024), with patch merging operations that progressively increase the number of feature channels while reducing spatial resolution.

During decoding, patch expanding operations are applied to progressively recover spatial resolution while reducing channel dimensionality. In contrast to the original Unet, skip connections between corresponding encoder and decoder stages are implemented via element-wise addition rather than feature concatenation.

VM-Unet was trained on an NVIDIA A800 GPU. Network parameters were optimized using the AdamW optimizer (Loshchilov and Hutter, 2017), with a learning rate of 0.001, $\beta_1 = 0.9$, $\beta_2 = 0.999$, $\varepsilon = 10^{-8}$, and a weight decay of 0.01. The batch size was set to 8, and training was conducted for 10 epochs. When training on Training Dataset C, which contained 40,863 pseudo GMH-IVH slices, only 1% of the images were randomly sampled per epoch to mitigate overfitting. In addition, the VSS blocks in both the encoder and decoder were initialized with pre-trained VMamba weights (vmamba_small_e238_ema) (Ruan and Xiang, 2024) to facilitate faster convergence.

**eAppendix 2. Detailed SAM Architecture and Training Procedures**

The Segment Anything Model (SAM) (Kirillov et al., 2023) is a versatile image segmentation framework trained on approximately 11 million natural images with more than 1 billion annotated masks, enabling strong zero-shot generalization across a wide range of segmentation tasks. However, SAM tends to underperform when segmenting medical images characterized by low contrast, ambiguous boundaries, complex morphology, or small target structures. To improve cross-domain generalization, several SAM fine-tuning strategies have been proposed to adapt pretrained representations to medical imaging applications. In this study, we fine-tuned an adapted SAM to enhance the accuracy of hemorrhage region segmentation in fetal brain MRI.

Specifically, we fine-tuned the SAMUS model (Lin et al., 2024), which preserves the core SAM architecture by retaining the original prompt encoder and mask decoder without modification. The image encoder was adapted by reducing the input matrix size from 1024 × 1024 to 256 × 256 to better match the spatial resolution commonly encountered in medical images. In addition, a convolutional neural network (CNN) branch and a cross-branch attention (CBA) mechanism were incorporated to facilitate cross-domain feature adaptation, enabling effective fine-tuning on the GMH-IVH dataset.

During training, SAMUS took both the image and a point prompt as inputs. Point prompts were generated by randomly sampling points within lesion regions based on the ground-truth segmentation masks. Model training was performed on an NVIDIA A800 GPU, with network parameters optimized using the Adam optimizer (Kingma, 2014). The optimizer was configured with a learning rate of $1 \times 10^{-4}$, $\beta_1 = 0.9$, $\beta_2 = 0.999$, and a weight decay of 0.01. A batch size of 8 was used, and training was conducted for 20 epochs. When training on Training Dataset C, only 0.5% of the images were randomly selected per epoch to reduce the risk of overfitting.

**eAppendix 3. Architectures and Training of Anomaly Detection Models**

In annotation-free lesion detection, unsupervised anomaly detection (UAD) (Bercea et al., 2023) can also be employed, in addition to the pseudo-anomaly generation method proposed in this study. These models detect lesion data by modeling the distribution of normal image features and identifying outliers. To compare our proposed method with representative UAD approaches, we trained three knowledge distillation–based anomaly detection models (Skip-TS (M. Liu et al., 2024), IKD (Cao et al., 2022), and RD4AD (Deng and Li, 2022)) using Training Dataset B.

IKD (Informative Knowledge Distillation) (Cao et al., 2022) is a UAD model designed to improve previous knowledge distillation approaches. It introduced two key innovations: a Context Similarity Loss, which helped the student network learn the structure of the normal data manifold, and Adaptive Hard Sample Mining, which focused training on informative hard samples. By distilling more informative knowledge from teacher to student, IKD mitigated overfitting and achieved state-of-the-art performance on anomaly segmentation benchmarks, particularly for complex object categories. HRNet-32 (Wang et al., 2020) was used as the backbone for both the student and teacher networks. The model was trained on an NVIDIA A800 GPU, with network parameters optimized using the AdamW optimizer (Loshchilov and Hutter, 2017) (learning rate = 0.00005). A batch size of 2 was used, and training was conducted for 5 epochs.

RD4AD (Reverse Distillation for Anomaly Detection) (Deng and Li, 2022) introduced a novel "reverse distillation" paradigm for UAD. RD4AD consisted of a pre-trained teacher encoder, a trainable one-class bottleneck embedding (OCBE) module, and a student decoder. The student learned to reconstruct the teacher's multi-scale representations from a compact embedding. Anomalies were detected by measuring discrepancies between the teacher's and student's features, with the heterogeneous encoder-decoder design and reverse knowledge flow improving anomaly discrimination. WideResNet50 (Zagoruyko and Komodakis, 2016) was used as the backbone for the teacher encoder, and a mirrored version of WideResNet50 was used for the student decoder. The model was trained on an NVIDIA A800 GPU, with parameters optimized using the Adam optimizer (Kingma, 2014) (learning rate = 0.005). A batch size of 16 was used, and training lasted 10 epochs.

Skip-TS (Teacher-Student Model with Skip Connections) (M. Liu et al., 2024) builds on RD4AD by introducing a direct reverse knowledge distillation paradigm for unsupervised anomaly detection in medical images. Skip-TS comprised a pre-trained teacher encoder and a randomly initialized student decoder connected via skip connections, enabling reconstruction of normal image representations at multiple scales. Anomalies were detected based on discrepancies between the teacher's and student's multi-scale features, and the heterogeneous architecture with direct feature transfer improved both anomaly discrimination and localization. Similar to RD4AD, WideResNet50 (Zagoruyko and Komodakis, 2016) was used for the teacher encoder and a mirrored WideResNet50 for the student decoder. The model was trained on an NVIDIA A800 GPU, with parameters optimized using the Adam optimizer (Kingma, 2014) (learning rate = 0.005), a batch size of 16, and 10 epochs of training.

**eAppendix 4. FreeHemoSeg Demonstration**

This video provides a comprehensive demonstration of FreeHemoSeg, a deep learning-based tool for automated detection and segmentation of fetal germinal matrix-intraventricular hemorrhage (GMH-IVH) on brain MRI. First, structural MRI data are imported to generate an AI-driven anomaly heatmap for GMH-IVH detection. The heatmap provides both case-level and slice-level anomaly scores and allows users to adjust visualization parameters, such as opacity and contrast. Second, the video highlights the tool's full brain tissue segmentation capability, including automatic volume quantification of structures such as the ventricles and white matter. Finally, the anomaly heatmap is thresholded to segment the lesion and compute hemorrhage volume, enabling quantitative clinical assessment.

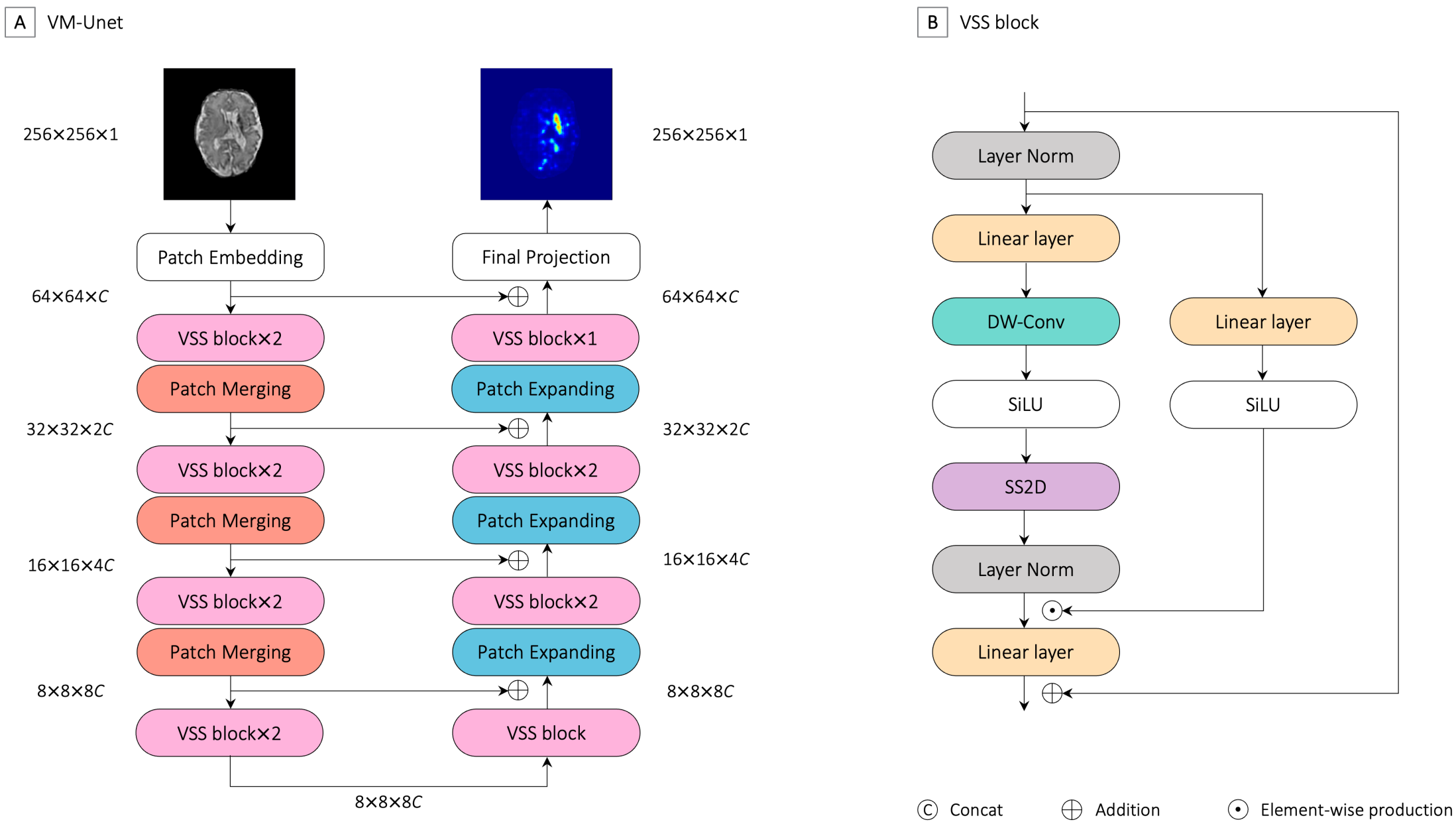

**eFigure 1. Detailed Architecture of VM-Unet.**

VM-Unet is an enhanced Unet architecture incorporating Vision Mamba, with the Visual State Space (VSS) block serving as its core component

A Importing Fetal Brain Volume Data

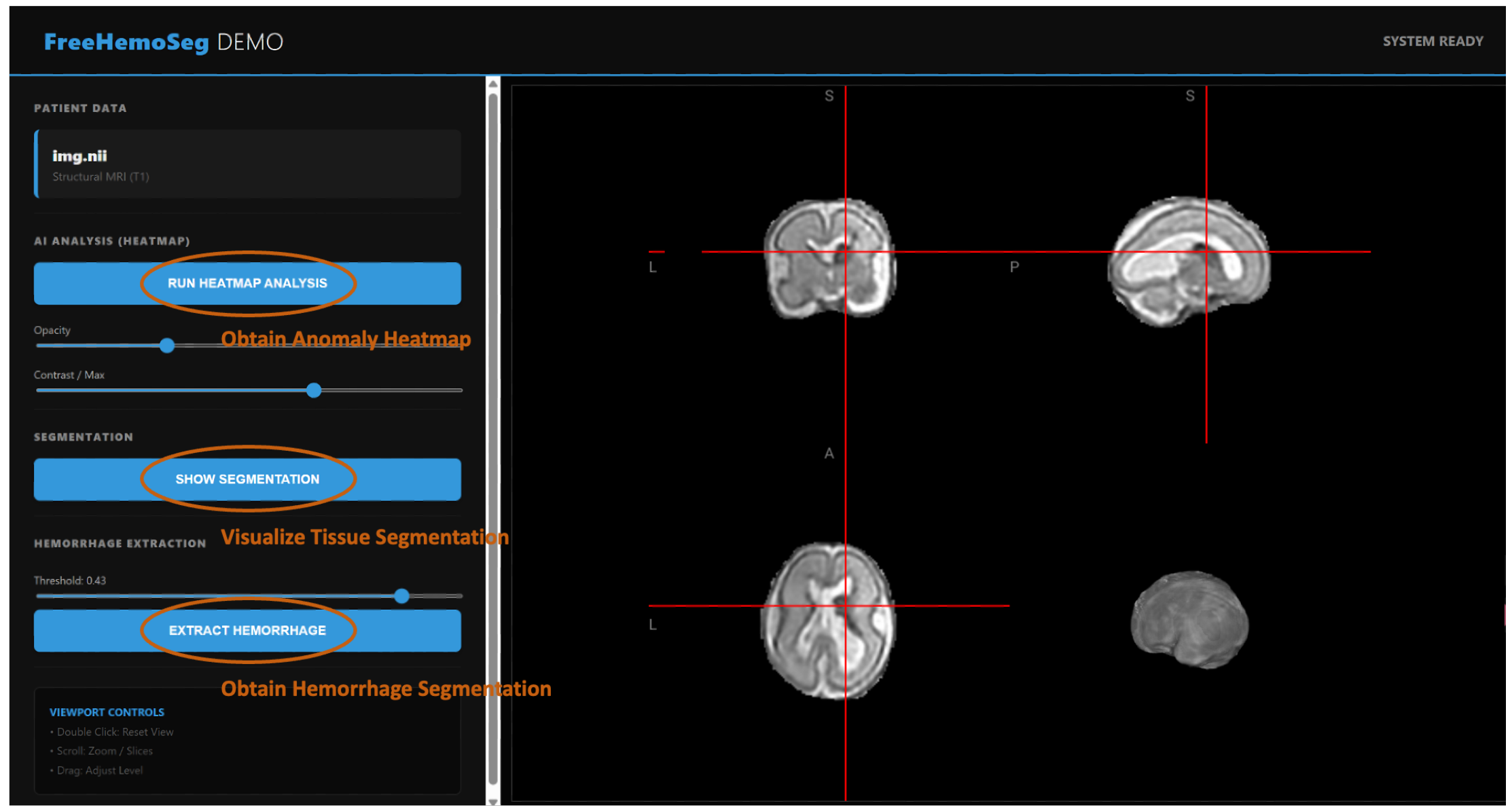

B Run FreeHemoSeg to Obtain Anomaly Heatmap

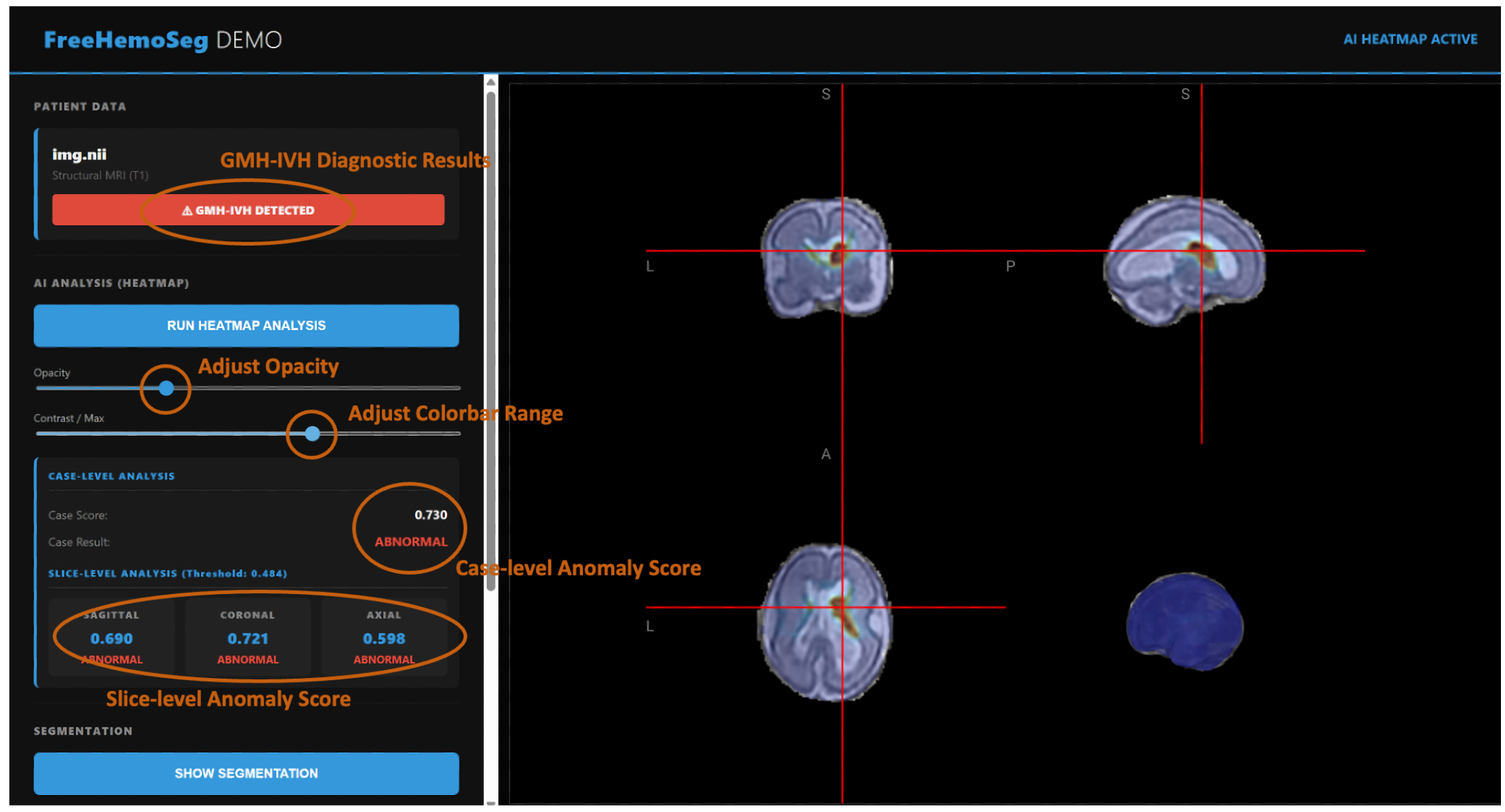

**eFigure 2. FreeHemoSeg Demonstration.**

The original fetal brain images and corresponding FreeHemoSeg output heatmaps are shown with comprehensive annotations, including case-level and slice-level anomaly scores, as well as guidance for adjusting visualization parameters.

C Visualize Tissue Segmentation Results

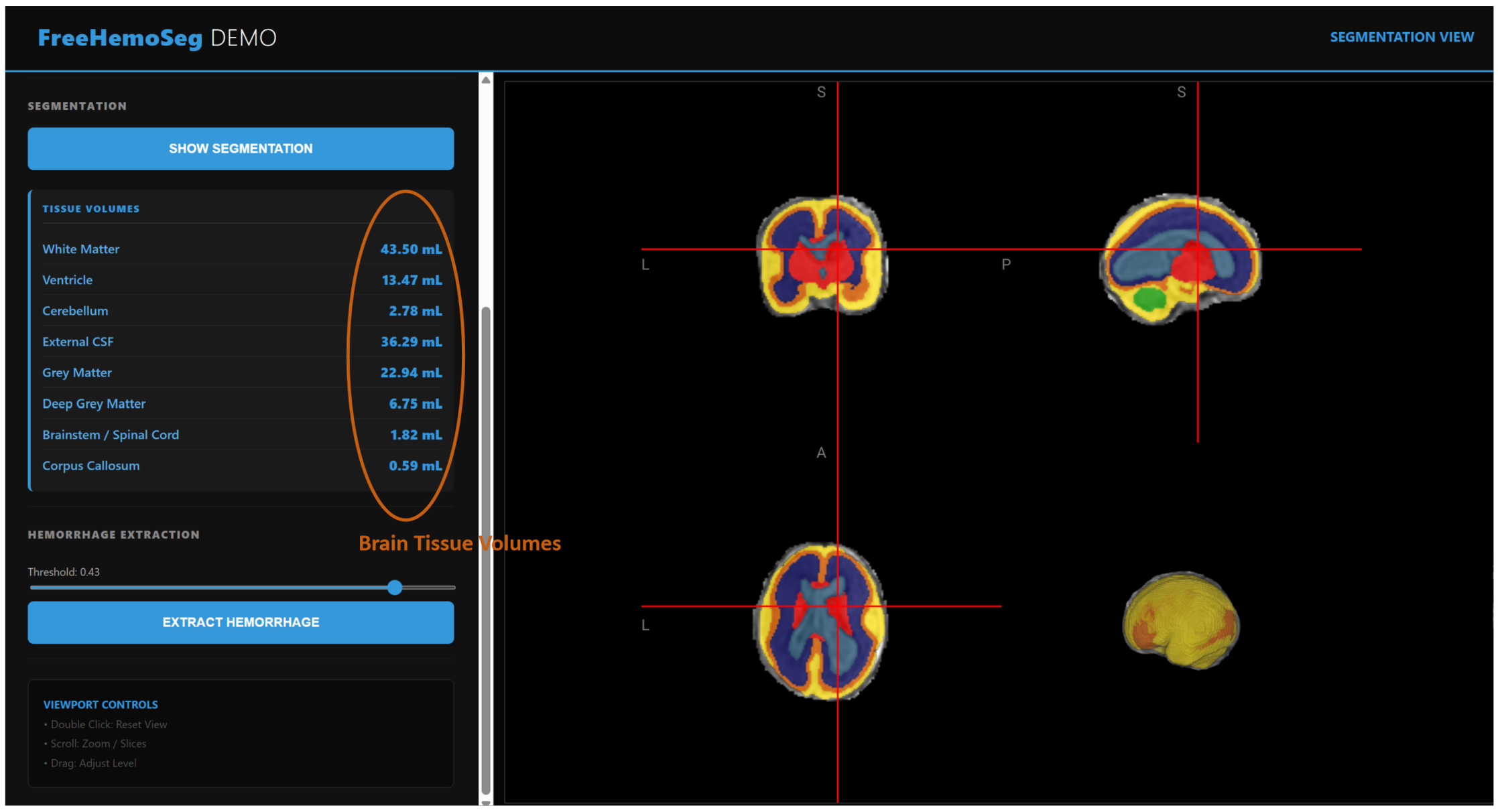

D Threshold Anomaly Heatmap to Obtain Hemorrhage Segmentation

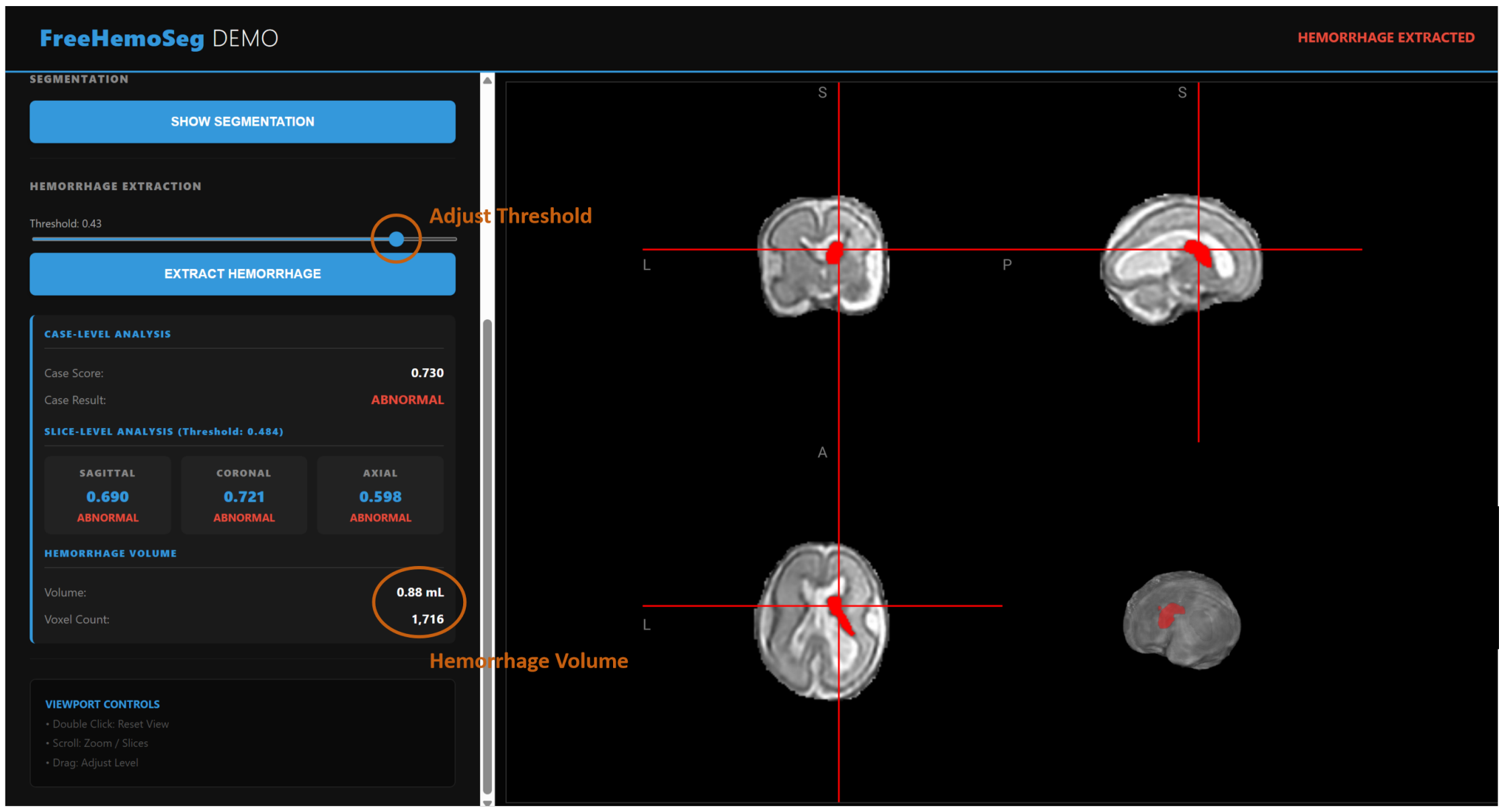

**eFigure 3. FreeHemoSeg Demonstration (Continued).**

Tissue segmentation maps and hemorrhage segmentation results are shown with comprehensive annotations, including brain tissue and hemorrhage volumes, along with guidance for adjusting visualization parameters.

**eReferences**

Bercea, C.I., Puyol-Antón, E., Wiestler, B., Rueckert, D., Schnabel, J.A., King, A.P., 2023. Bias in Unsupervised Anomaly Detection in Brain MRI.

Cao, Y., Wan, Q., Shen, W., Gao, L., 2022. Informative knowledge distillation for image anomaly segmentation. Knowl.-Based Syst. 248, 108846. https://doi.org/10.1016/j.knosys.2022.108846

Deng, H., Li, X., 2022. Anomaly Detection via Reverse Distillation from One-Class Embedding, in: 2022 IEEE/CVF Conference on Computer Vision and Pattern Recognition (CVPR). Presented at the 2022 IEEE/CVF Conference on Computer Vision and Pattern Recognition (CVPR), IEEE, New Orleans, LA, USA, pp. 9727–9736. https://doi.org/10.1109/CVPR52688.2022.00951

Kingma, D.P., 2014. Adam: A method for stochastic optimization. ArXiv Prepr. ArXiv14126980.

Kirillov, A., Mintun, E., Ravi, N., Mao, H., Rolland, C., Gustafson, L., Xiao, T., Whitehead, S., Berg, A.C., Lo, W.-Y., 2023. Segment anything, in: Proceedings of the IEEE/CVF International Conference on Computer Vision. pp. 4015–4026.

Lin, X., Xiang, Y., Yu, L., Yan, Z., 2024. Beyond Adapting SAM: Towards End-to-End Ultrasound Image Segmentation via Auto Prompting, in: Linguraru, M.G., Dou, Q., Feragen, A., Giannarou, S., Glocker, B., Lekadir, K., Schnabel, J.A. (Eds.), Medical Image Computing and Computer Assisted Intervention – MICCAI 2024. Springer Nature Switzerland, Cham, pp. 24–34.

Liu, M., Jiao, Y., Lu, J., Chen, H., 2024. Anomaly Detection for Medical Images Using Teacher-Student Model with Skip Connections and Multi-scale Anomaly Consistency. IEEE Trans. Instrum. Meas. 1–1. https://doi.org/10.1109/TIM.2024.3406792

Liu, Y., Tian, Y., Zhao, Y., Yu, H., Xie, L., Wang, Y., Ye, Q., Liu, Yunfan, 2024. VMamba: Visual State Space Model. ArXiv abs/2401.10166.

Loshchilov, I., Hutter, F., 2017. Decoupled weight decay regularization. ArXiv Prepr. ArXiv171105101.

Ruan, J., Xiang, S., 2024. VM-UNet: Vision Mamba UNet for Medical Image Segmentation.

Wang, J., Sun, K., Cheng, T., Jiang, B., Deng, C., Zhao, Y., Liu, D., Mu, Y., Tan, M., Wang, X., 2020. Deep high-resolution representation learning for visual recognition. IEEE Trans. Pattern Anal. Mach. Intell. 43, 3349–3364.

Zagoruyko, S., Komodakis, N., 2016. Wide Residual Networks. ArXiv Prepr. ArXiv160507146.